\pdfoutput=1

\documentclass{article}
\usepackage[utf8]{inputenc}

\usepackage{graphicx}
\usepackage{amsmath}
\usepackage[version=4]{mhchem}
\usepackage{siunitx}
\usepackage{longtable,tabularx}
\usepackage{subcaption}
\usepackage[shortlabels]{enumitem}
\usepackage{hyperref}

\usepackage{grffile}
\usepackage{nth}

\graphicspath{{./figures/}}

\usepackage[usenames,dvipsnames,svgnames,table]{xcolor}
\usepackage[subfigure]{tocloft}
\usepackage{changes}
\usepackage{xifthen}
\usepackage{wasysym}
\usepackage{algorithm}
\usepackage{algpseudocode}
\usepackage{amsopn}
\usepackage{bm}
\usepackage{multicol}

\makeatletter

\algnewcommand\NewParameterLine{%
	\newline \hspace*{\algorithmicindent} \hspace*{\algorithmicindent}
}

\algnewcommand\LeftComment[1]{%
	\Statex \vspace{0.5\baselineskip}\hspace{\ALG@thistlm}$\triangleright$  #1\hfill %
}

\algnewcommand\FirstLeftComment[1]{%
	\Statex \vspace{0.5\baselineskip}\hspace{\algorithmicindent}\hspace{\ALG@thistlm}$\triangleright$  #1\hfill %
}

\algnewcommand\FirstLeftCommentCont[1]{%
	\Statex \hspace{\algorithmicindent}\hspace{\ALG@thistlm}\phantom{$\triangleright$}  #1\hfill %
}
\algnewcommand\MultiLineState{%
	\Statex \hspace{\algorithmicindent}\hspace{\ALG@thistlm}%
}
\makeatother

\newcounter{exampleCounter}
\refstepcounter{exampleCounter}

\renewcommand{\vec}[1]{\mathbf{#1}}

\newcommand{\mat}[1]{\mathbf{#1}}

\providecommand{\url}[1]{\texttt{#1}}

\newcommand{\norm}[2][]
{
	\ensuremath{\left\| #2\right\|
		\ifthenelse{\isempty{#1}}
		{}
		{_{{#1}}}}\xspace
}

\newcommand{\gradient}[1]
{\ensuremath{\nabla#1}\xspace
}

\newcounter{todoListCounter}

\newcommand{\theTodoListCounter}
{
  \arabic{todoListCounter}
}

\newcommand{\notDone}
{\textcolor{Sepia}{$\Square$}}

\newcommand{\listTodoName}
{ \ifthenelse{\theTodoListCounter >1}
  {
    There are \textcolor{Sepia}{\theTodoListCounter} todo's
  }
  {
    There is \textcolor{Sepia}{\theTodoListCounter} todo
  }
}

\newlistof{todo}{tmp}{\listTodoName}

\providecommand{\newtodo}[2]
{
  \refstepcounter{todo}
  \noindent \rule{\linewidth}{1pt} \\
  \textbf{\large \theTodoListCounter} \textcolor{Sepia}{ TODO}: {\sc #2} \hfill \raisebox{-1ex}{\Huge #1}\\
  \rule{\linewidth}{1pt} \\
  \vspace{-1.0\baselineskip}
  \addcontentsline{tmp}{todo}{{\large #1} \textcolor{Sepia}{TODO \numberline{\thetodo:}}\textcolor{black}{#2}}
}

\newcommand{\listOfTodo}
{
  \ifthenelse{\theTodoListCounter > 0}
  {
    \clearpage
    \listoftodo
  }
  {
  }
}

\newcommand{\eval}[2]
{\ensuremath{#1\mathopen{}\left(#2\right)\mathclose{}}}

\newcommand{\Functional}[1]
{E_{#1}}

\newcommand{\format}[1]
{\texttt{#1}}

\usepackage{xspace}

\algrenewcommand\algorithmicindent{1.0em}

\newcommand{\distortionFunction}
{\ensuremath{\eta\xspace}}

\newcommand{\qualityFunction}
{\ensuremath{\mathrm{q}\xspace}}

\newcommand{\mapName}
{\ensuremath{\bm{\phi}^*\xspace}}

\newcommand{\mapNameB}
{\ensuremath{\bm{\phi}\xspace}}

\newcommand{\elementName}
{\ensuremath{e\xspace}}

\newcommand{\domainName}
{\ensuremath{\Omega\xspace}}

\newcommand{\meshName}
{\ensuremath{\mathcal{M}\xspace}}

\newcommand{\pCoord}
{\ensuremath{\vec{x}}\xspace}

\newcommand{\iCoord}
{\ensuremath{\vec{y}}\xspace}

\newcommand{\distortion}[2][]
{\ifthenelse{\isempty{#2}}
	{\ensuremath{\distortionFunction_{#1}}}
	{\ensuremath{\distortionFunction_{#1}(#2)}}\xspace
}

\newcommand{\modDistortion}[2][]
{\ifthenelse{\isempty{#2}}
	{\ensuremath{\distortionFunction_{0}}}
	{\ensuremath{\distortionFunction_{0}(#2)}}\xspace
}

\newcommand{\quality}[2][]
{\ifthenelse{\isempty{#2}}
	{\ensuremath{\qualityFunction_{#1}}}
	{\ensuremath{\qualityFunction_{#1}(#2)}}\xspace
}

\newcommand{\modQuality}[2][]
{\ifthenelse{\isempty{#2}}
	{\ensuremath{\qualityFunction_{#1}^*}}
	{\ensuremath{\qualityFunction_{#1}^*(#2)}}\xspace
}

\newcommand{\objectiveFunction}[2][]
{\ifthenelse{\isempty{#2}}
	{\ensuremath{f_{#1}}}
	{\ensuremath{f_{#1}(#2)}}\xspace
}

\newcommand{\h}[1]
{\ifthenelse{\isempty{#1}}
	{\ensuremath{\determinant_\delta}}
	{\ensuremath{\determinant_\delta(#1)}}\xspace
}

\newcommand{\jacobian}[1][]
{\ifthenelse{\isempty{#1}}
	{\ensuremath{\mat{S}}}
	{\ensuremath{\mat{S}(#1)}}\xspace
}

\newcommand{\map}[2][]
{\ensuremath{\mapName
\ifthenelse{\isempty{#2}}
	{{}}
	{_{#2}}
\ifthenelse{\isempty{#1}}
	{{}}
	{(#1)}}\xspace
}

\newcommand{\mapB}[2][]
{\ensuremath{\mapNameB
\ifthenelse{\isempty{#2}}
	{{}}
	{_{#2}}
\ifthenelse{\isempty{#1}}
	{{}}
	{(#1)}}\xspace
}

\newcommand{\mapD}[2][]
{\ensuremath{\mapNameB_h
    \ifthenelse{\isempty{#2}}
    {{}}
    {_{#2}}
    \ifthenelse{\isempty{#1}}
    {{}}
    {(#1)}}\xspace
}

\newcommand{\determinant}[1][]
{\ifthenelse{\isempty{#1}}
	{\ensuremath{\sigma}}
	{\ensuremath{\sigma(#1)}}\xspace
}

\newcommand{\element}[1][]
{\ifthenelse{\isempty{#1}}
	{\ensuremath{\elementName}}
	{\ensuremath{\elementName_{#1}}}\xspace
}

\newcommand{\domain}[2][]
{\ensuremath{\domainName
	\ifthenelse{\isempty{#1}}
	{{}}
	{^{#1}}_{#2}}\xspace
}

\newcommand{\Gradient}[3][]
{\ensuremath{
	\nabla
	\ifthenelse{\isempty{#1}}
		{}
		{_{#1}}
	#2
	\ifthenelse{\isempty{#3}}
		{}
		{(#3)}
}\xspace
}

\newcommand{\Jacobian}[2][]
{\ifthenelse{\isempty{#1}}
	{\ensuremath{\textrm{\textbf{D}}#2}}
	{\ensuremath{\textrm{\textbf{D}}#2(#1)}}\xspace
}

\newcommand{\mesh}[2][]
{\ensuremath{\meshName^{{#1}}_{{#2}}}\xspace}

\newcommand{\projection}[2]
{\eval{\Pi_{#1}}{#2}}

\DeclareMathOperator*{\argmin}{arg\,min}

\newcommand{\node}[1]
{\ifthenelse{\isempty{#1}}
	{\ensuremath{v}}
	{\ensuremath{\vec x_{#1}}}\xspace
}

\newcommand{\scalarProduct}[3][]
{\ensuremath{\langle #2, #3\rangle
\ifthenelse{\isempty{#1}}
	{}
	{_{#1}}}\xspace
}

\newcommand{\trace}
{\ensuremath{\boldsymbol T}\xspace}

\newcommand{\virtualSurface}{\ensuremath{\mathcal{S}}\xspace}
\newcommand{\virtualCurve}{\ensuremath{\mathcal{C}}\xspace}

\title{Generation of Curved Meshes \\ for the High-Lift Common Research Model}

\author{
	Eloi Ruiz-Giron\'es
	\footnote{Computer Applications in Science and Engineering, \texttt{eloi.ruizgirones@bsc.es}.}
	and
	Xevi Roca
	\footnote{Computer Applications in Science and Engineering, \texttt{xevi.roca@bsc.es}}
}

\date{Barcelona Supercomputing Center - BSC, Barcelona, Spain, 08034}

\begin{document}

\maketitle

\begin{abstract}
We answer the questions of the high-order technology focus group (HO-TFG) about the mesh generation for the high-lift common research model of the \nth{4} high-lift prediction workshop. The HO-TFG seeks answers about the feasibility of generating meshes for complex geometries, and how to measure the quality of different aspects of the mesh. To answer these questions, we first generate several curved meshes and then, we analyze different aspects of the curved mesh and perform a visual inspection.  The main bottleneck of our curving methodology is the preparation of curving-friendly inputs, a process that can take several days for complex geometries. Our distributed parallel implementation executed with 768 processors is able to curve the presented meshes in minutes.
\end{abstract}

\section{Introduction}
\label{sec:introduction}

The objective of the \nth{4} High-Lift Prediction Workshop (HLPW4) \cite{rumsey2022:HLPW4_overview} is to predict the high-lift flows around a complex geometry using state-of-the-art simulation tools. The geometry considered in the workshop is the high-lift common research model (CRM-HL) \cite{lacy2020:crm-hl_definition, evans2020:crm-hl_test}, a standardized aircraft in a high-lift configuration. The workshop is organized as a collaboration of the international community, in which participants are grouped in several technology focus groups that deal with different aspects of the simulation process. In particular, the high-order technology focus group (HO-TFG) \cite{hlpw4_ho-tfg} posed questions about the high-order approximation of both the geometry and the numerical solution.

In this work, we answer the high-order mesh generation questions of the HO-TFG:
\begin{enumerate}[1.]
	\item Can 3D curved meshes be generated for the CRM-HL?
	\item What mesh quality metrics are used to evaluate high order meshes?
	\item How well do the curved meshes conform to the actual geometry?
\end{enumerate}
In summary, the HO-TFG seeks answers about the feasibility of generating meshes for complex geometries, and how to measure the quality of different aspects of the mesh. The answer of the first question is primordial since a curved mesh is required to perform a simulation with unstructured high-order methods. The answer to the second question provides tools to analyze the mesh quality \emph{a priori}. In this manner, it is not necessary to perform an expensive simulation to test a curved mesh. Finally, the answer to the third question checks the geometric accuracy of the curved meshes. This is important to faithfully simulate the appropriate immersed object.

To answer these questions, we first present the methodology that we have used to generate the curved high meshes for the CRM-HL and then, we analyze the resulting meshes. We divide our curving process in three steps. In the pre-process step, we create a virtual geometry model, generate a linear mesh, and assign boundary marks. Then, in the curving step, we solve a non-linear problem in a distributed parallel environment to obtain a high-order mesh that approximates the virtual model. Finally, in the post-process step, we check several aspects of the curved mesh, perform a visual inspection, and export the mesh in the \format{cgns} parallel file format.

To analyze the final meshes, we check the validity of the mesh, the quality of the elements, and the geometric accuracy of the mesh. These are key aspects of the mesh that lead to high-quality simulations. In particular, the mesh validity is required to perform a simulation. The element quality is important since low-quality elements may introduce spurious artifacts to the numerical solution and may hamper the convergence of the linear and non-linear solvers. Finally, the geometric accuracy of the mesh is necessary to faithfully simulate the appropriate immersed object. Moreover, we also perform a visual inspection of the mesh to check the smoothness of the boundary triangles. In particular, we need to avoid oscillations in the interior of the boundary triangles, and large differences between the normal vector of adjacent triangles.

Our main bottleneck to obtain a curved mesh is the preparation of curving-friendly inputs. We should avoid tangent entities in the geometry definition that lead to invalid meshes. Moreover, the linear mesh has to be feasible to curve and therefore, we have to avoid some element configurations that hamper or complicate the curving process. Finally, we need an appropriate resolution to approximate the target geometry with the required accuracy. To satisfy these requirements, we may spend several days adjusting both the virtual model and the linear mesh. On the contrary, our curving method executed in parallel with 768 processors spends several minutes to curve a single mesh composed of millions of tetrahedra.

The rest of the paper is structured as follows. In Section \ref{sec:parallelMeshCurving}, we detail our mesh curving methodology. In Section \ref{sec:examples} we show the generated meshes and analyze them. In Section \ref{sec:answers} we answer the HO-TFG questions. Finally, in Section \ref{sec:concludingRemarks} we present the concluding remarks.

\section{Parallel mesh curving}
\label{sec:parallelMeshCurving}

We perform the curving process in three steps. In the pre-process step, we create a virtual geometry model, generate an initial linear mesh, and prepare all the required input files. In the curving step, we solve a non-linear problem to obtain a curved mesh. Finally, in the post-process step, we perform a visual inspection of the resulting curved mesh and we analyze several quality metrics.

\subsection{Pre-process}

\subsubsection{Geometry repair and virtual geometry model}

\begin{figure*}[t!]
	\centering
	\hfill
	\begin{subfigure}[b]{0.27\textwidth}
		\includegraphics[width=\textwidth]{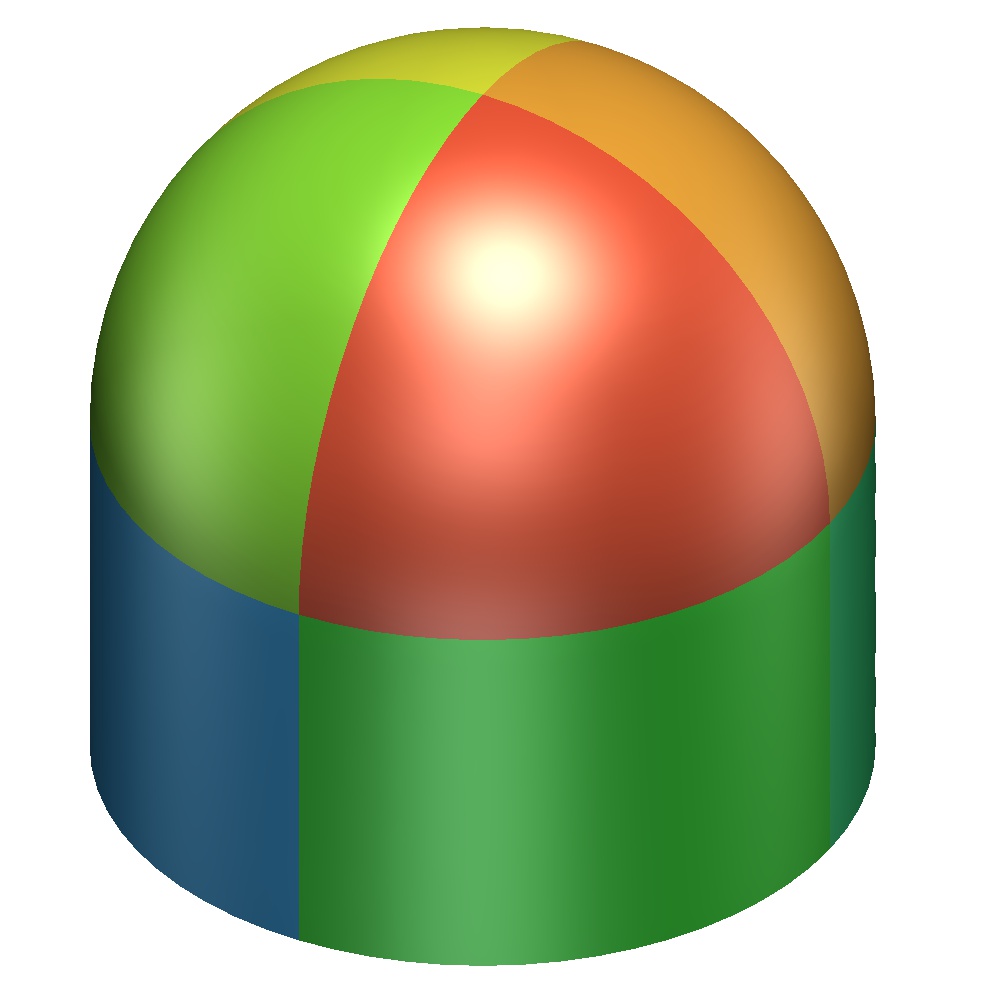}
		\caption{}
		\label{fig:bullet}
	\end{subfigure}
	\hfill
	\begin{subfigure}[b]{0.27\textwidth}
		\includegraphics[width=\textwidth]{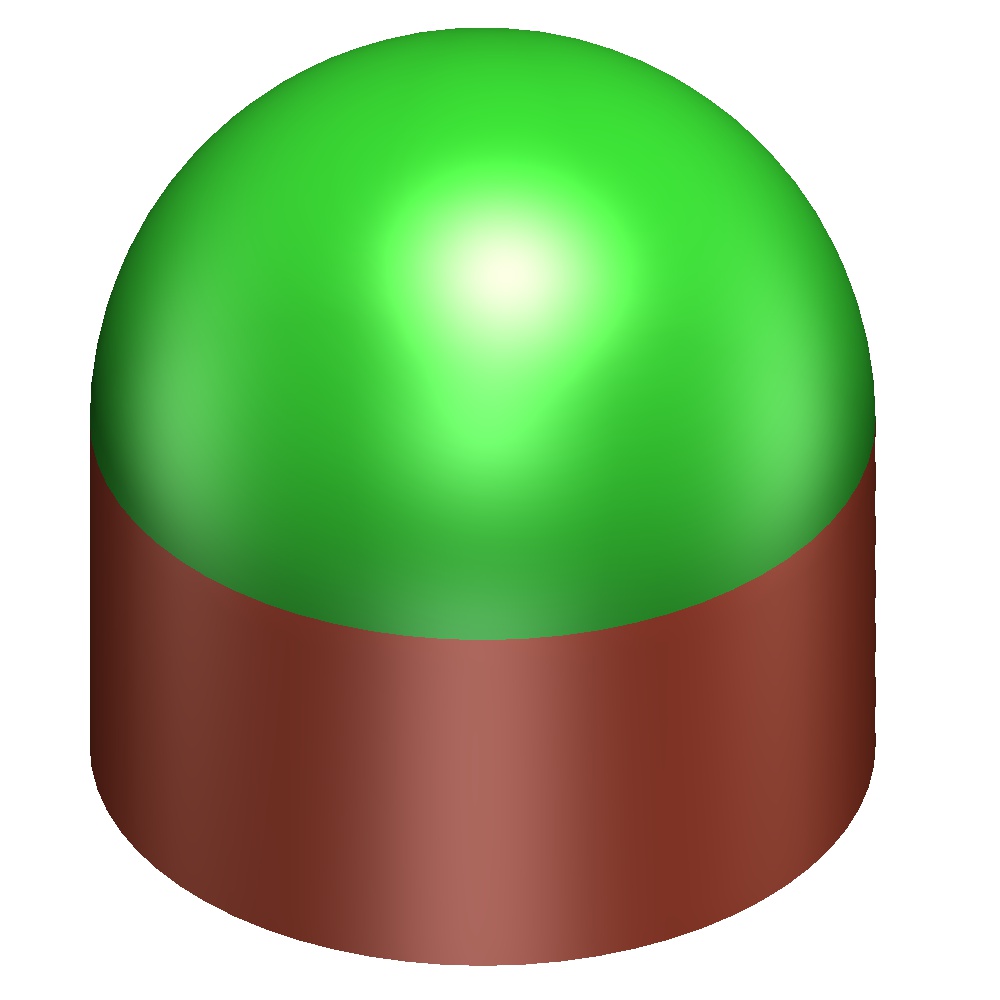}
		\caption{}
		\label{fig:bullet_virtual}
	\end{subfigure}
	\hfill\hspace{0cm}
	\caption{CAD model colored according to the different boundary entities using:
		(a) the original CAD surfaces; and
		(b) a virtual model.}
	\label{fig:virtualGeometry}
\end{figure*}

To generate an initial linear mesh, we need a clean and water-tight CAD model. Thus, the first step is to repair the CAD model and close the gaps between surfaces. Once the geometric model is closed, we group the surfaces and curves of the model into virtual entities using the virtual engine of Pointwise \cite{pointwise}. The virtual entities allow decoupling the topology of the geometric model from the topology of the mesh \cite{ruiz2018:industrialHO}. Therefore, we can generate meshes of higher quality. The main idea is to group the surfaces in which the normal between adjacent surfaces is continuous. In this manner, the final curved mesh will approximate the virtual model with smooth elements.

Figure \ref{fig:virtualGeometry} shows a CAD model colored according to the different surfaces. The original model is composed of nine surfaces, that correspond to four surfaces for the upper hemisphere, four surfaces for the  cylinder, and a cap surface at the bottom, see Figure \ref{fig:bullet}. Nevertheless, we can group the surfaces of the hemisphere into a single and smooth virtual entity. We repeat the same process to create a smooth virtual entity for the cylinder. The final virtual model is composed of three virtual entities, see Figure \ref{fig:bullet_virtual}. Note that the final mesh will not inherit the topology of the artificial decomposition of the original surfaces.

\subsubsection{Linear mesh generation}

The next step of the process is the generation of an initial linear mesh. To perform this step, we use the Pointwise program \cite{pointwise}. The linear mesh needs to satisfy several requirements according to the simulation accuracy, the geometric approximation and the mesh curving feasibility. We need to ensure that the linear mesh contains elements of the desired shape and size according to the simulation requirements. For instance, we need a high resolution around the immersed object, and a low resolution in the far-field. In addition, we need to generate a boundary layer around the aircraft with the desired wall distance and growing rate.

To approximate the target virtual geometry with sufficient accuracy, the boundary mesh needs enough resolution. To this end, we generate small elements around curves and surfaces with high curvature, and larger elements around the entities with low curvature. In this manner, we will obtain a curved mesh with enough geometric accuracy. Otherwise, the normal vector between adjacent boundary triangles may not be similar, and this will affect the accuracy of the numerical solution.

We also need to avoid configurations of the elements that precludes the generation of a curved mesh. If these configurations are present, even if the linear mesh is valid, the resulting curved mesh will contain invalid elements. The first configuration is a boundary triangle with two edges along tangent curves. In this case, the curved triangle will contain two tangent edges that will lead to a null Jacobian. The second configuration is a tetrahedron with two triangles on tangent surfaces. Similarly as in the previous case, the curved tetrahedron will contain two tangent faces, and a null Jacobian along the edge.

During the curving process, we need to project the high-order boundary nodes onto the target virtual entities. To facilitate this process, we assign an integer identifier to the boundary triangles according to the virtual surface they approximate. We perform this action in an automatic manner using the \format{glyph} scripting language of Pointwise. In the script, we iterate through the triangular meshes that approximate each virtual surface, and create a boundary mark for each one. When exporting the linear mesh, each triangle in the boundary contains an integer identifier.

\subsubsection{Creating the input files}

Using Pointwise, we store the mesh file and the boundary conditions in the \format{gmsh} format. Nevertheless, the \format{gmsh} format cannot be read in a parallel distributed manner. Reading the mesh sequentially in a distributed environment introduces a bottleneck since there is only one processor reading the mesh. This is an important issue when dealing with a large number of processors. For this reason, we create a new mesh file in the \format{hdf5} format. The \format{hdf5} format allows us to read and write to a file in a distributed manner. Thus, we avoid the bottleneck of reading the initial mesh.

To read the \format{gmsh} file and store the \format{hdf5} file, we use a python script using the FEniCS library \cite{alnaes:fenics}. The resulting \format{hdf5} file contains the linear mesh and the boundary marks in the FEniCS format. The FEniCS library allows reading the file in a distributed manner and we avoid the bottleneck of a sequential read.

We store the virtual model that we have created using the pointwise format \format{nmb}. In our code, we  read the \format{nmb} files using a python wrapper of the geode lite library. The CAD file is necessary to use the geometric model to project points onto surfaces. All processors read the whole CAD model and, in this manner, we are able to project points onto surfaces in our distributed parallel environment. The drawback of this approach is that all processors contain all the CAD model. Nevertheless, in our applications, the memory footprint of the CAD is negligible compared to the Hessian matrix contribution in each processor.

Finally, we have a file that classifies each boundary mark into the far field, the symmetry plane and the immersed object. To this end, we create a numpy \format{npz} file that contains three arrays of integer ids corresponding to these boundary mark. We use the classification of boundary ids for three purposes. First, when we output the final curved mesh in \format{cgns} format, we export the boundary marks according to this classification. Second, when curving the linear mesh, we fix the far field boundary in order to simplify the curving process. Finally, when computing the geometric accuracy of the curved mesh, we only take into account the curved triangles that approximate the immersed object.

\subsection{Mesh curving problem}

\begin{figure*}[h!]
	\centering
	\hfill
	\includegraphics[width=0.45\textwidth]{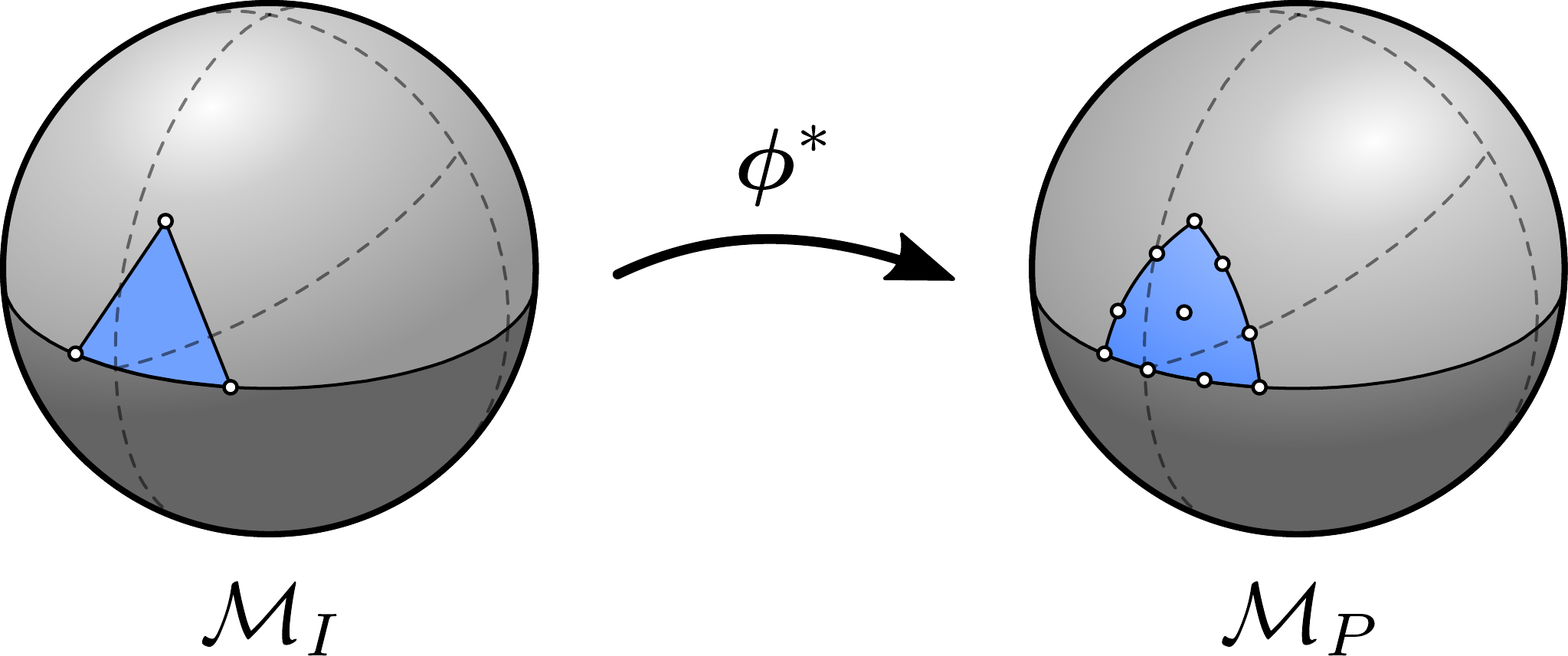}
	\hfill\hspace{0cm}
	\caption{Initial linear mesh and optimized curved mesh approximating a virtual geometry.}
	\label{fig:meshCurving}
\end{figure*}

\subsubsection{Mesh curving formulation}

Given an initial linear mesh, \mesh[]{I}, we want to characterize a curved high-order one, \mesh[]{P}, in terms of a diffeomorphism \map{} \cite{gargallo2015:tetOptimization, ruiz2016Hierarchical}. The optimal diffeomorphism presents optimal point-wise distortion, and satisfies a prescribed boundary condition, see Figure \ref{fig:meshCurving}. That is, \map{} is the minimizer of
\begin{align}
	&\min_{\mapNameB \in \mathcal V} \eval{E}{\mapNameB} = \norm[\mesh{I}]{M\mapNameB}^2 \nonumber\\
	&\text{subject to:} \nonumber \\
	&\trace \mapNameB = \eval{\vec{g}_D}{\trace \mapB{}},
	\label{eqn:constrained}
\end{align}
where
\[
	\norm[\mesh{I}]{f}^2 = \int_{\mesh{I}} f^2 \text{d}\domain{},
\]
\trace is the trace operator, $\eval{\vec{g}_D}{\trace \mapB{}}$ is a non-linear Dirichlet boundary condition on $\partial\mesh{I}$ that depends on the values of \mapB{},  and
\[
	\eval{M\mapB{}}{\iCoord} =
	\eval{\eta}{\Jacobian{\eval{\mapB{}}{\iCoord}}}
	= 
	\frac{\norm[]{\eval{\Jacobian{\mapB{}}}{\iCoord}}^2}{n \eval{\determinant_0}{\eval{\Jacobian{\mapB{}}}{\iCoord}}^{2/n}}
\]
is a regularized point-wise distortion measure \cite{gargallo2015:tetOptimization} defined in terms of the shape distortion measure for linear simplices \cite{knupp:algebraicQuality}, where \norm{\cdot} is the Frobenius norm for matrices, and
\begin{equation}
	\determinant_0=\frac{1}{2}\left ( \determinant + |\determinant| \right),
	\label{eqn:pModDistortion}
\end{equation}
being \eval{\determinant}{\cdot} the determinant function. The regularized distortion measure takes a value of infinity when the determinant is negative or equal to zero, and takes finite values when the determinant is positive.

The non-linear boundary condition allows using a geometric model in the mesh curving process. The proposed formulation allows different kinds of boundary conditions by modifying the function $\vec{g}_D$. For instance, we can use a fixed boundary condition by setting a constant value. Moreover, we can slide the nodes along the CAD entities by introducing the parametric coordinates of the boundary nodes into the curving problem. To generate the meshes for the High-Lift Prediction Workshop, we have used a boundary condition that allows integrating a virtual model as
\begin{equation}
	\eval{\vec g_D}{\trace \mapB[]{}} = 
	\sum_{i=1}^{N_b} \projection{}{\pCoord_i} N_i^b,
	\label{eqn:phiProjection}
\end{equation}
where $\pCoord_i$ are the coordinates of the mesh nodes, $N_b$ is the number of boundary nodes, $\{N^b\}_{i=1,\ldots,N_b}$ is a Lagrangian basis of shape functions that are $\mathcal{C}^0-$continuous between adjacent boundary faces, and \projection{}{\cdot} is a geometric projection operator onto the CAD model. The boundary condition can be interpreted as an interpolation of the geometric model, in which the interpolation points are the projection of the boundary nodes. This boundary condition is non-linear because it depends on the projection of the boundary nodes.

To define the projection operator $\Pi$, we differentiate two cases. The first one is the projection onto virtual surfaces, and the second one is the projection onto virtual curves. We define the projection of a point \pCoord onto a virtual surface \virtualSurface as
\begin{equation}
	\projection{\virtualSurface}{\pCoord} =
	\argmin_{\vec y \in \virtualSurface} \norm{\pCoord - \vec y}.
	\label{eqn:surfaceProjection}
\end{equation}
Note that the point $\vec y$ belongs to the virtual surface \virtualSurface and therefore, to actually perform the projection computation, we loop over all the surfaces contained in \virtualSurface and perform the point projection. The result of the operation is the projection point with minimum distance among all the surfaces contained in \virtualSurface.

To deal with gaps between curves and its adjacent surfaces, we avoid projecting points to virtual curves. Instead, we define the curve projection operator using the projection to the adjacent virtual surfaces of the virtual curves. Let $\virtualSurface_1$ and $\virtualSurface_2$ be the adjacent virtual surfaces of a virtual curve \virtualCurve. We define the projection of a point \pCoord onto \virtualCurve as
\begin{equation}
	\projection{\virtualCurve}{\pCoord} = \frac{1}{2}\Big(
	\projection{\virtualSurface_1}{\projection{\virtualSurface_2}{\pCoord}} +
	\projection{\virtualSurface_2}{\projection{\virtualSurface_1}{\pCoord}}
	\Big)
	\label{eqn:curveProjection}
\end{equation}

We perform the projection against virtual curves in the proposed manner in order to deal with the gaps between adjacent surfaces and curves. When using the proposed curve projection, the projected point is an average of the projection points onto the two surfaces. In this manner, we allow the curved mesh to traverse the gap between adjacent surfaces in a smooth manner.

To read a virtual geometry from a file and to perform the point projection in our code, we have implemented a python wrapper of Pointwise's geode lite library \cite{geode}. The geode library allows us to iterate through the virtual entities contained in the geometry model, and to iterate through the entities contained in the virtual entities. Moreover, the library is able to project points onto CAD surfaces. Using these functionalities, we implemented in the wrapper the point projection procedure described in Equation \eqref{eqn:surfaceProjection}.

\subsubsection{Mesh curving solver}

To solve the constrained optimization problem in \eqref{eqn:constrained}, we use a penalty approach, see \cite{ruiz2018:industrialHO}, in which we introduce the boundary constraint into the objective function in a weak sense as follows
\begin{equation}
	\min_{\mapB{} \in \mathcal V}
	\eval{\Functional{\mu}}{\mapB{}} = 
	\frac{\eval{\Functional{}}{\mapB{}}}{\norm[\mesh{I}]{1}^2} + 
	\mu \frac{\norm[{\partial \mesh{I}}]{\trace \mapB{} - \eval{\vec{g}_D}{\trace \mapB{}}}^2}{\norm[\partial \mesh{I}]{1}^2},
	\label{eqn:unconstrained}
\end{equation}
where
	\[
	\norm[\partial\mesh{I}]{f}^2 = \int_{\partial\mesh{I}} f^2 \text{d}\domain{},
	\]
$\mu$ is a penalty parameter that enforces the validity of the constraint when it tends to infinity. We have introduced the measures of the initial mesh and its boundary in order to balance the two contributions of the new functional.

The main idea is to solve several unconstrained optimization problems with increasing penalty parameter in order to enforce the boundary condition. Nevertheless, the boundary condition depends on the actual solution of the problem. Thus, we apply a fixed-point iteration as
\begin{equation*}
	\vec g_D^{k} = \eval{\vec g_D}{\trace \mapB[]{}^k}, \qquad
	\mapB{}^{k+1} = \argmin_{\mapB{} \in \mathcal V} \eval{\Functional{\mu}}{\mapB{}; \vec g_D^k},
\end{equation*}
being $k$ the $k$-th iteration of the proposed fixed-point solver.

We optimize each non-linear problem of the proposed penalty method using a backtracking line-search method in which the advancing direction is computed using Newton's method and the step-length is set using the Armijo's rule, see \cite{nocedal:optimization} for more details.

The convergence criterion of the penalty method is defined in terms of the gradient of the functional, and the value of the constraint as:
\[
\norm[\partial{\mesh{I}}]{\mapB{}^k - \trace \mapB{}^k} / \norm[\partial{\mesh{I}}]{1}< \varepsilon^*,
\qquad
\norm[\infty]{\gradient{\eval{\Functional{\mu}}{\mapB{}^k; \vec g_D^k}}} < \omega^*.
\]
That is, we terminate the penalty method when both the boundary error and the residual are small enough. In our applications, we set $\omega^* = 10^{-8}$ and $\varepsilon^* = 10^{-12} \ell_c$, being $\ell_c$ a characteristic length of the model.

\subsubsection{Solver improvements}

We propose four improvements to reduce the computational cost and memory requirements of generating a curved mesh in references \cite{ruiz2019:pContinuation, ruiz2022:parallelCurving}. The proposed improvements are essential to curve larger meshes without increasing neither the waiting time or the required computational resources. As a consequence, we also reduce the energy consumption of generating a curved mesh.

To reduce the computational cost, we propose three main ingredients. The first one is a $p$-continuation technique \cite{ruiz2019:pContinuation}. Instead of directly computing the optimal mesh for a given polynomial degree, we iterate through the polynomial degrees and optimize them. The initial condition for each polynomial degree is the optimized mesh of the previous one. In this manner, most of the iterations of the linear and non-linear problems are performed in the lower polynomial degrees. Since lower polynomial degrees lead to problems with less degrees of freedom, these iterations are less costly than the ones performed for higher polynomial degrees.

The second ingredient to improve the computational cost is to reduce the number of non-linear problems solved during the penalty method. In the penalty method, we increase the penalty parameter to enforce the boundary condition. For each value of the penalty parameter, we solve a non-linear problem and increase the penalty parameter. We iterate this process until the solver achieves convergence. Nevertheless, as shown in \cite{ruiz2022:parallelCurving}, we can predict the value of the penalty parameter to converge the curving process. Thus, instead of incrementing the penalty parameter by a constant factor, we can set the optimal penalty parameter to converge the method. Since we solve a non-linear problem for each value of the penalty parameter, we end up solving less non-linear problems.

The third ingredient to reduce the computational cost is to reduce the number of iterations of the linear solvers. To this end, we adapt the tolerance to converge the linear problems \cite{ruiz2022:parallelCurving}. The main idea is that it is not necessary to solve all the linear systems with the same tolerance. In the first non-linear iterations of the curving process, it is possible to select a loose tolerance to solve the linear systems since we are \emph{far} to the optimal solution. In the last iterations of the process, we want to obtain the quadratic convergence of Newton's method and therefore, we need to use a tight tolerance in order to accurately solve the linear system. Using appropriate tolerances, we can avoid unnecessary iterations of the linear solver and therefore, increase the computational efficiency of the code.

Moreover, we reduce the memory requirements of solving a linear system three times \cite{ruiz2022:parallelCurving}. Therefore, we are able to solve problems with three times more elements without increasing the computational requirements. The main idea is to solve the linear problems using a matrix-free gmres method, and using a reduced-memory pre-conditioner. The proposed pre-conditioner is a block-based SOR method, in which each block is associated with a dimension of the problem. When applying the pre-conditioner, we only need to store the three diagonal blocks. Therefore, the memory requirements are divided by three.

\subsection{Post-process}

\subsubsection{Quality metrics of the curved high-order mesh}

Once we have curved the mesh, we check the element quality and the geometric accuracy. To this end, we first ensure that the final mesh contains valid elements. In our curving process, we detect invalid configurations at the integration points. To further ensure that the mesh is valid, we increase the resolution of the integration points and compute two element quality measures. Thus, we check the mesh validity, but we also check the mesh quality. This is an important ingredient since the mesh quality influences several aspects of a simulation process. Low quality meshes may increase the condition number of the finite element matrices and therefore, increment the difficulty of solving the linear and non-linear problems. Moreover, low quality meshes may introduce spurious artifacts in the numerical solution. For this reason, it is key not only to obtain a valid mesh, but also a high quality mesh.

The first element quality metric is the relative shape quality measure of the curved elements defined in \cite{gargallo2015:tetOptimization} as:
\[
	q_{\element[P]}^{S} = \frac{1}{\eta_{\element[P]}},
	\quad \text{where} \quad
	\eta_{\element[P]} = \left(
	\frac
	{\displaystyle \int_{\element[I]{}} (M\mapB[]{})^2\ \text{d}\domain{}}
	{\displaystyle \int_{\element[I]{}} 1 \ \text{d}\domain{}}
	\right)^{1/2}.
\]
The shape quality measure is equal to zero for inverted elements, and equal to one for ideal elements. The second quality measure is the scaled Jacobian, defined as:
\[
	q_{\element[P]}^{SJ} =
	\frac{\displaystyle \inf_{\pCoord \in \element[I]} \eval{\determinant{}}{\Jacobian{\mapB{}}}}
	{\displaystyle \sup_{\pCoord \in \element[I]} \eval{\determinant{}}{\Jacobian{\mapB{}}}}.
\]
The scaled Jacobian quality measures how curved is an element. It is equal to one when the maping \mapB{} is affine, and it is equal or less than zero when the element is tangled.

To check how accurately the curved mesh, \mesh{P}, approximates the geometric model, \domain{}, we compute three different geometric accuracy measures. The first one is
\[
	\eval{SC}{\mesh{P},\domain{}} =
	\frac{\displaystyle \int_{\mesh{B}} \norm{\vec x - \eval{\Pi}{\vec x}} \ \text{d}\Gamma}
	{\displaystyle \int_{\mesh{B}} 1 \ \text{d}\Gamma}
	= \quad
	\frac{\displaystyle \sum_{\virtualSurface_i \in \domain{}} \int_{\mesh{\virtualSurface_i}} \norm{\vec x - \projection{\virtualSurface_i}{\pCoord}} \ \text{d}\Gamma}
	{\displaystyle \int_{\mesh{B}} 1 \ \text{d}\Gamma},
\]
where \mesh{B} is the boundary of physical curved mesh, \mesh{\virtualSurface} is the curved triangular mesh that approximates the virtual surface \virtualSurface, \norm{\cdot} is the Euclidean norm of vectors, and \projection{\virtualSurface}{\cdot} is the projection operator discussed in Equation \eqref{eqn:surfaceProjection}. This geometric accuracy measure is the average distance between the curved mesh and the target geometry.

The second geometric accuracy measure is similar to the disparity measure \cite{ruiz2021:disparity}
\[
	\eval{d_2}{\mesh{P},\domain{}} = \left(
	\frac{\displaystyle \int_{\mesh{B}} \norm{\vec x - \eval{\Pi}{\vec x}}^2 \ \text{d}\Gamma}
	{\displaystyle \int_{\mesh{B}} 1 \ \text{d}\Gamma}
	\right)^{1/2} = \quad
	\left(
	\frac{\displaystyle \sum_{\virtualSurface_i \in \domain{}} \int_{\mesh{\virtualSurface_i}} \norm{\vec x - \projection{\virtualSurface_i}{\pCoord}}^2 \ \text{d}\Gamma}
	{\displaystyle \int_{\mesh{B}} 1 \ \text{d}\Gamma}
	\right)^{1/2}.
\]
In this case, the geometric accuracy is a $\mathcal L_2$-average of the distances between the curved mesh and the target domain. Thus, it penalizes the larger distances.

The last geometric accuracy measure is
\[
	\eval{d_\infty}{\mesh{P},\domain{}} = \sup_{\pCoord \in \mesh{B}} \norm{\pCoord - \eval{\Pi}{\pCoord}},
\]
which only takes into account the largest point-wise distance between the curved mesh and the target geometry.

\subsubsection{Visual inspection of the curved mesh}

We perform a visual inspection to check different aspects of the curved mesh. To this end, we export the curved mesh in the high-order parallel format of Paraview \cite{ahrens2005:paraview}. The exported mesh contains the initial linear mesh, the curved mesh as a high-order function over the linear mesh, and different quality measures for each element.

During the visual inspection, we locate the lower quality elements since these are the elements that negatively affect a simulation. Usually, there are two issues that lead to low element quality. The first one is related to bad geometry approximations. In this case, the elements are excessively curved when trying to approximate the target geometry. To solve this issue, we need to generate smaller elements in those areas. The second issue are constraints that the geometric model imposes on the mesh that lead to  poor configurations of elements, even when a virtual geometry engine is used. For instance, the CAD model may contain thin regions smaller than the element size or tangent curves.

The smoothness of the boundary mesh is an important factor for simulation purposes. Therefore, we check that the curved elements do not contain oscillations during the visual inspection. The oscillations in the boundary mesh are due to virtual surfaces with not enough geometric continuity between its composing surfaces. Moreover, we also check that the normal vector between elements is similar. To visually inspect the smoothness of the mesh, we check the continuity and smoothness of the specular highlights in Paraview.

\subsubsection{Export to \format{cgns} format}

In our python program, we store the curved mesh in a \format{hdf5} file using the FEniCS format. Nevertheless, to improve the exchange of curved meshes, we convert the final mesh to the \format{cgns} format. To do it so, we have implemented a python wrapper of the \format{cgns} library. Accordingly, we have developed a python script that reads the mesh in the FEniCS format and, using the \format{cgns} library wrapper, exports a \format{cgns} file. Currently, our mesh file converter accepts tetrahedral meshes up to polynomial degree four.

To export the curved mesh, we first reorder the nodes of each element according to the \format{cgns} format. Then, using the wrapper, we create an empty \format{cgns} file. In this file, we first write the nodes, and then the high-order elements. Finally, we extract the boundary triangles that approximate the far field, the aircraft and the symmetry plane. We export these triangles in the \format{cgns} file using the boundary condition tags \format{farfieldBC}, \format{wallBC}, and \format{symmetryBC}, respectively.

\subsection{Implementation details and used software}

We have generated the virtual model and the initial linear mesh using Pointwise. We have implemented our mesh curving solver in Python \cite{python} using the FEniCS \cite{alnaes:fenics}. Moreover, we also use FEniCS to read and write the mesh files in the parallel format \format{hdf5}. We solve the linear systems with the petsc4py \cite{petsc4py} libraries. To read the virtual model file and to project the nodes onto the virtual surfaces, we use the geode library \cite{geode} interfaced with a python wrapper using swig \cite{swig}. We store the final mesh in the \format{cgns} format with the cgns library interfaced with a python wrapper using pybind11 \cite{pybind11}.

We perform the mesh visualization using Paraview 5.5.2 in parallel in the MareNostrum4 super-computer. We have used the high-order mesh visualization implementation of Paraview that subdivides each element in a given number of sub-elements. Note that the mesh partition to perform the visualization does not need to coincide with the one used in the optimization. In general, for visualization purposes, less cores are needed since no global matrices are assembled and no linear systems are solved.

\section{Generation of curved meshes for the high-lift common research model}
\label{sec:examples}

We show the different steps that we performed to generate the curved meshes for the high-lift common research model. Specifically, we show the virtual model and the curved meshes, as well as the used quality metrics to analyze the meshes. Finally, we present the computational time to obtain the curved meshes.

\begin{figure*}[t!]
	\centering
	\hfill
	\begin{subfigure}[b]{0.475\textwidth}
		\includegraphics[width=\textwidth]{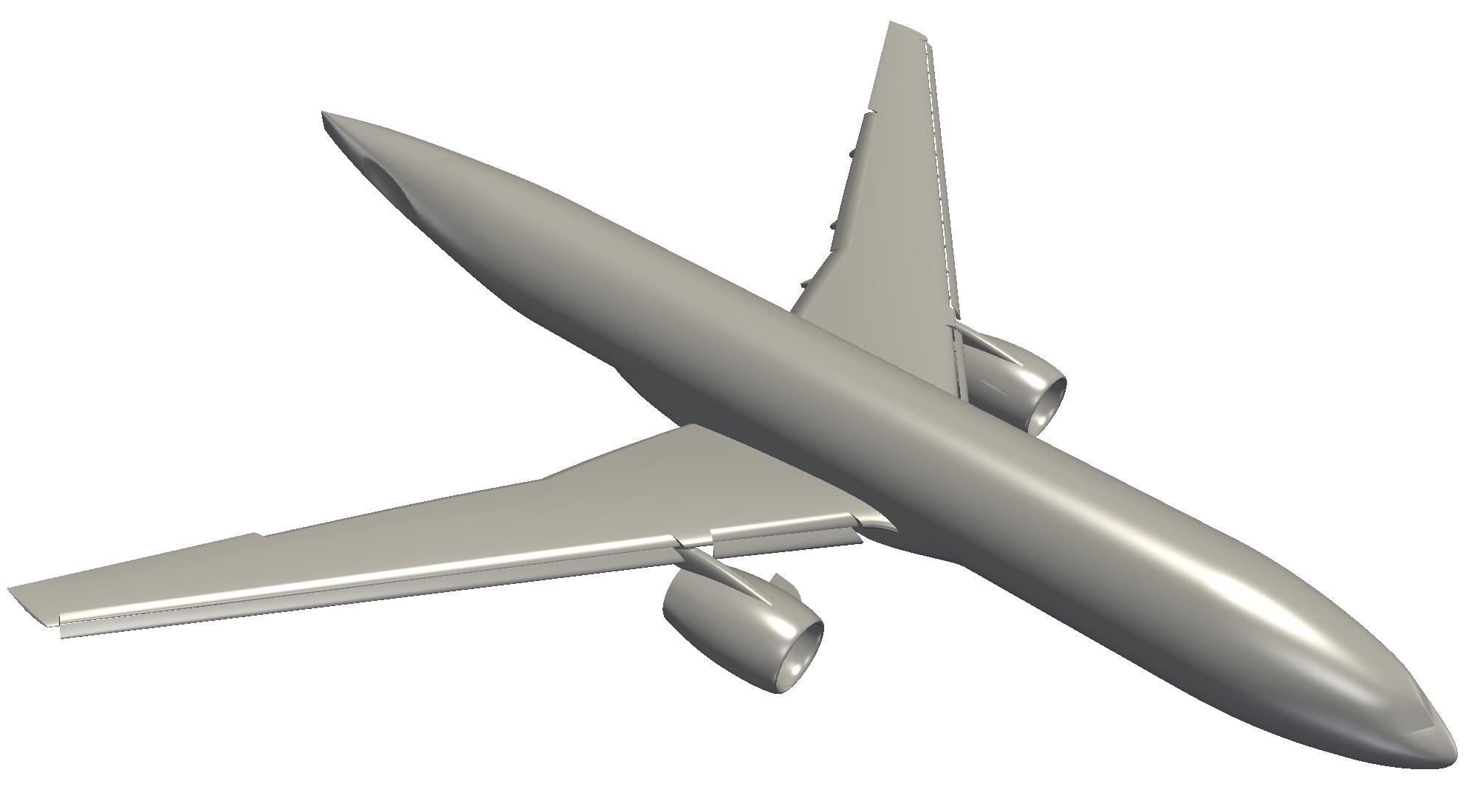}
		\caption{}
		\label{fig:crm_geometry_top}
	\end{subfigure}
	\hfill
	\begin{subfigure}[b]{0.475\textwidth}
		\includegraphics[width=\textwidth]{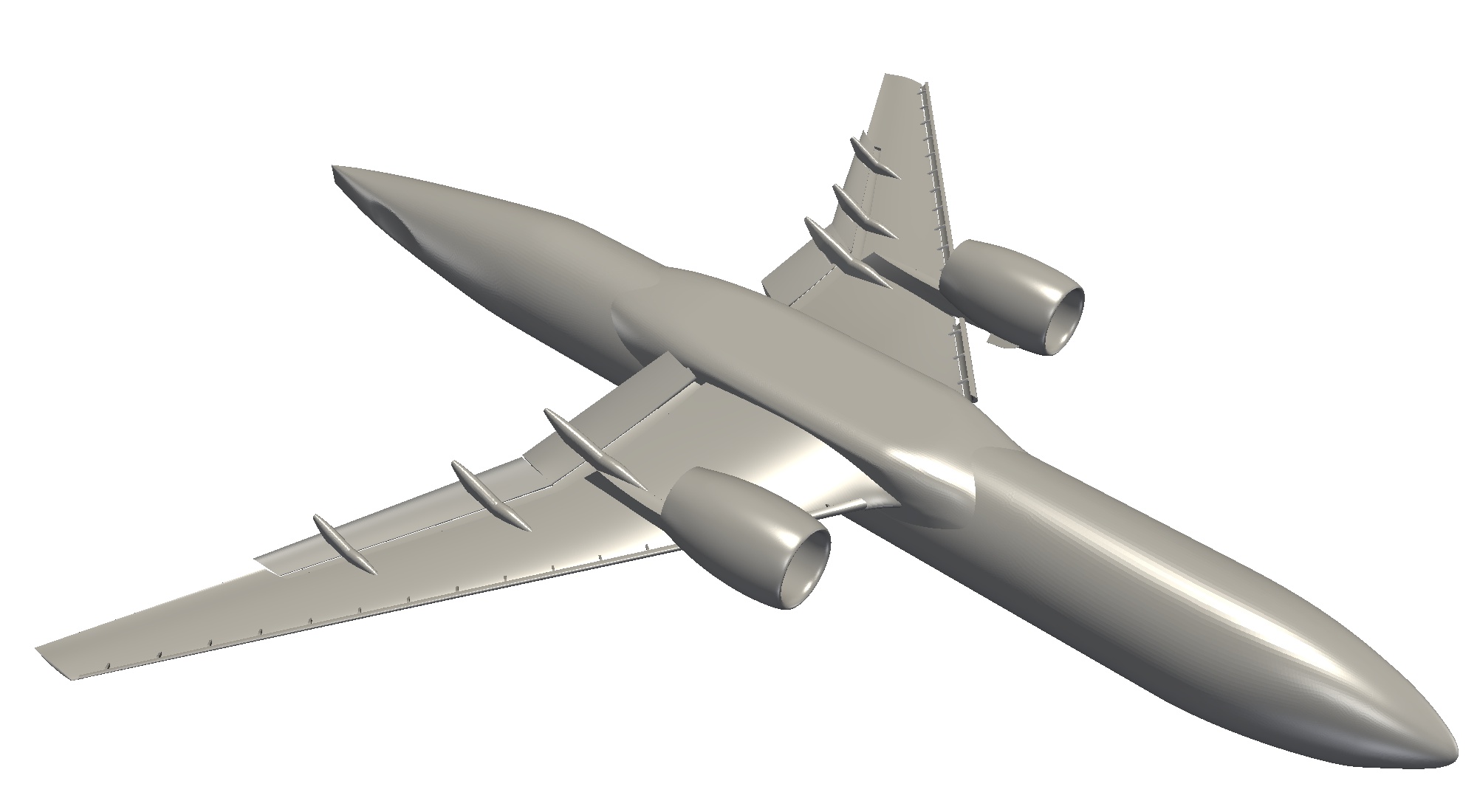}
		\caption{}
		\label{fig:crm_geometry_bottom}
	\end{subfigure}
	\hfill\hspace{0cm}
	\caption{Geometry of the high-lift common research model:
		(a) top view; and
		(b) bottom view.}
	\label{fig:crm_geometry}
\end{figure*}

\begin{figure*}[h!]
	\centering
	\hfill
	\begin{subfigure}[b]{0.475\textwidth}
		\includegraphics[width=\textwidth]{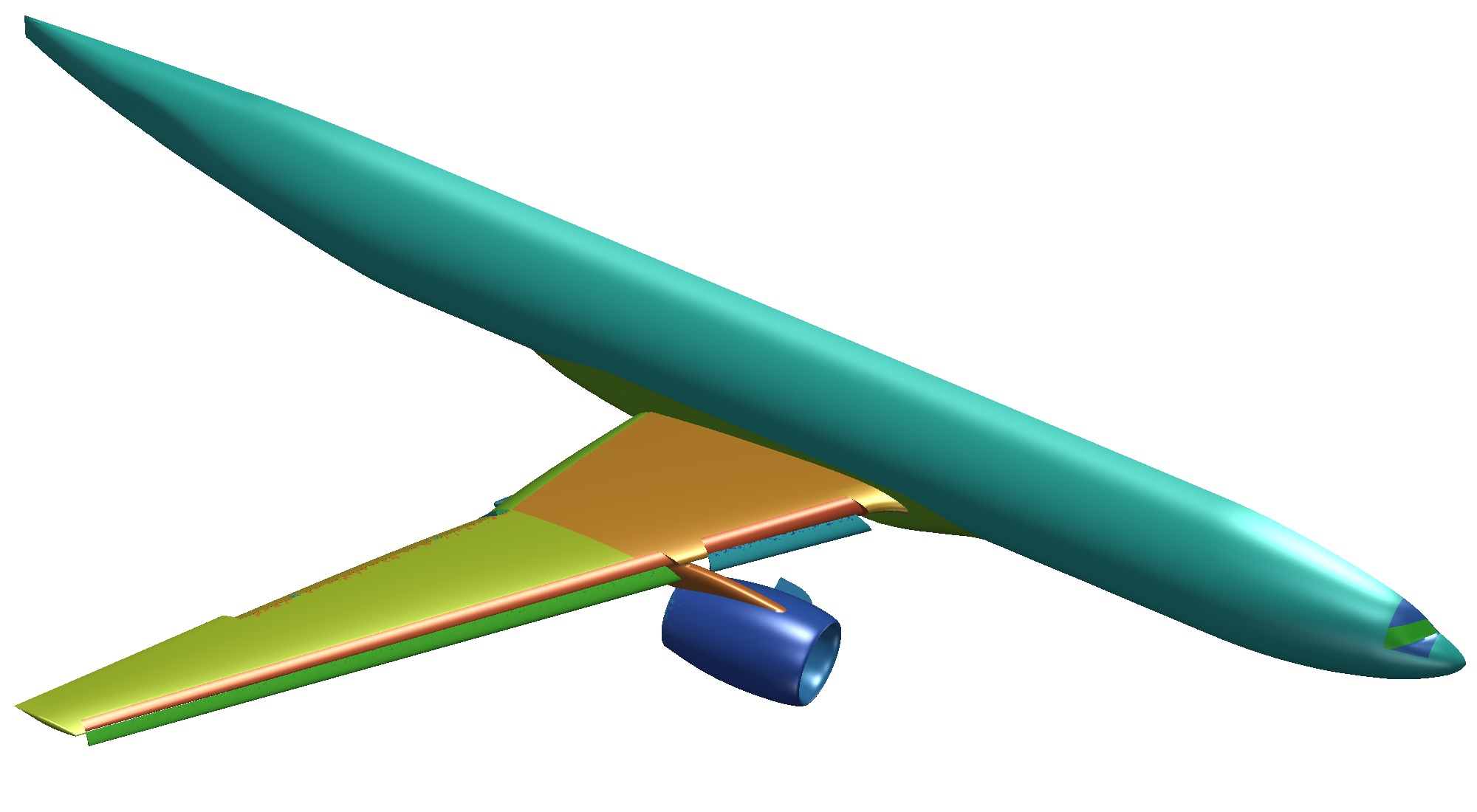}
		\caption{}
		\label{fig:crm_surfaces_top}
	\end{subfigure}
	\hfill
	\begin{subfigure}[b]{0.475\textwidth}
		\includegraphics[width=\textwidth]{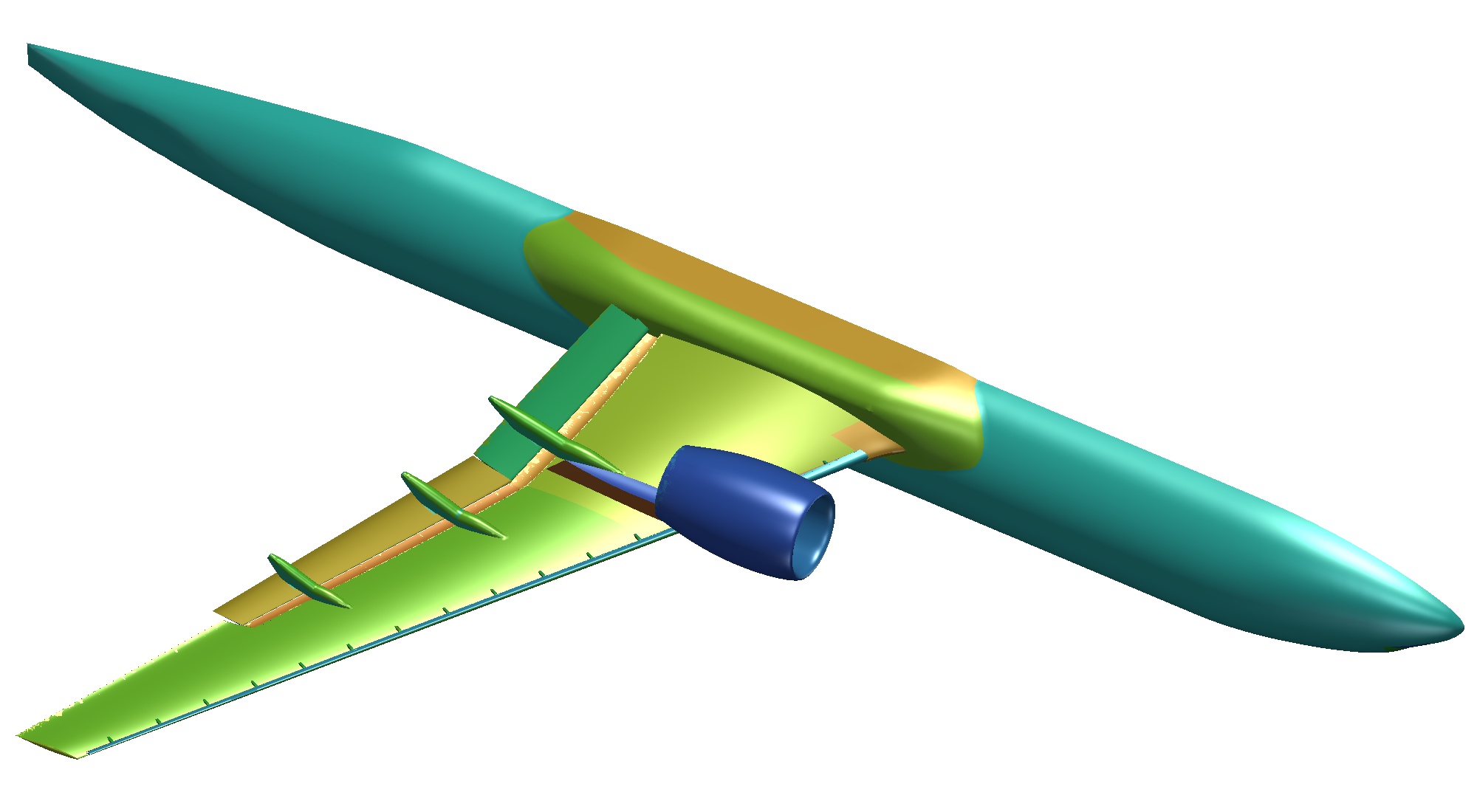}
		\caption{}
		\label{fig:crm_surfaces_bottom}
	\end{subfigure}
	\hfill\hspace{0cm}
		\caption{Virtual model used to curve the meshes for the high-lift common research aircraft:
		(a) top view; and
		(b) bottom view.}
	\label{fig:crm_surfaces}
\end{figure*}

\subsection{Virtual geometry model and curved mesh generation}

The full geometry of the aircraft, see Figure \ref{fig:crm_geometry}, is symmetric and therefore, we only need to mesh half of the geometry. Using the geometry, we create a virtual model by joining surfaces according to the normal continuity between them, see Figure \ref{fig:crm_surfaces}. In this manner, we substitute small surfaces by larger ones and therefore, we can generate larger elements that do not follow the topology of the original model. It is important to create virtual surfaces with enough continuity between its surfaces. Otherwise, the curved mesh may contain oscillations that hamper the simulation. The initial model contains 415 surfaces and the virtual model contains 215 virtual surfaces. 

With the virtual model, we have generated meshes of polynomial degree two and three, with different boundary layer configurations. Specifically, we have used an inviscid configuration without boundary layer, and three different boundary layers with a growing rate of $1.5$ and $Y+ = 800, 200, 100$.

\subsection{Visual inspection of the curved meshes}

\begin{figure*}[h!]
	\centering
	\hfill
	\begin{subfigure}[b]{0.90\textwidth}
		\includegraphics[width=\textwidth]{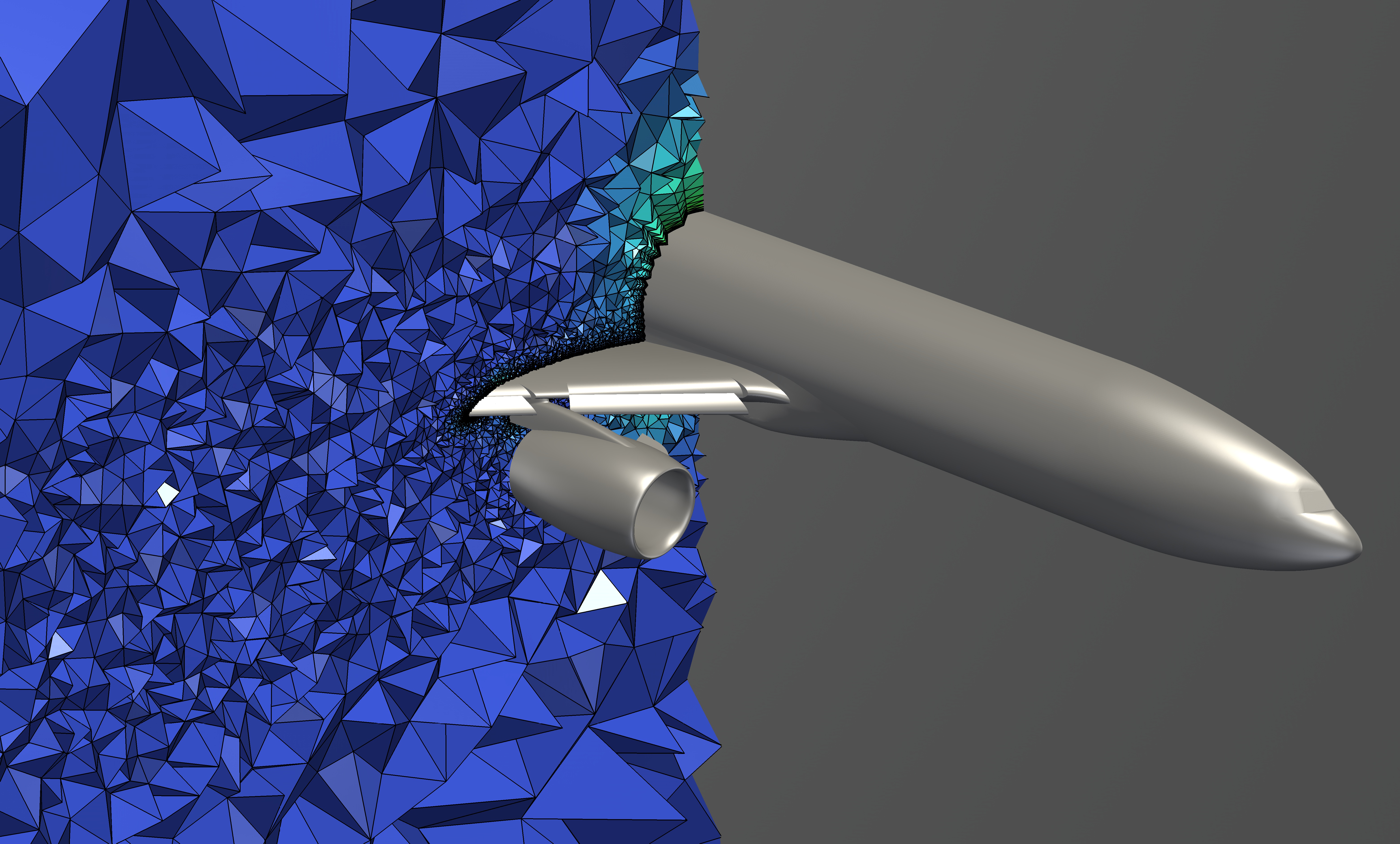}
		\caption{}
		\label{fig:CRM_Y100_P2_Cut}
	\end{subfigure}
	\hfill\hspace{0cm}
	\\
	\hfill
	\begin{subfigure}[b]{0.90\textwidth}
		\includegraphics[width=\textwidth]{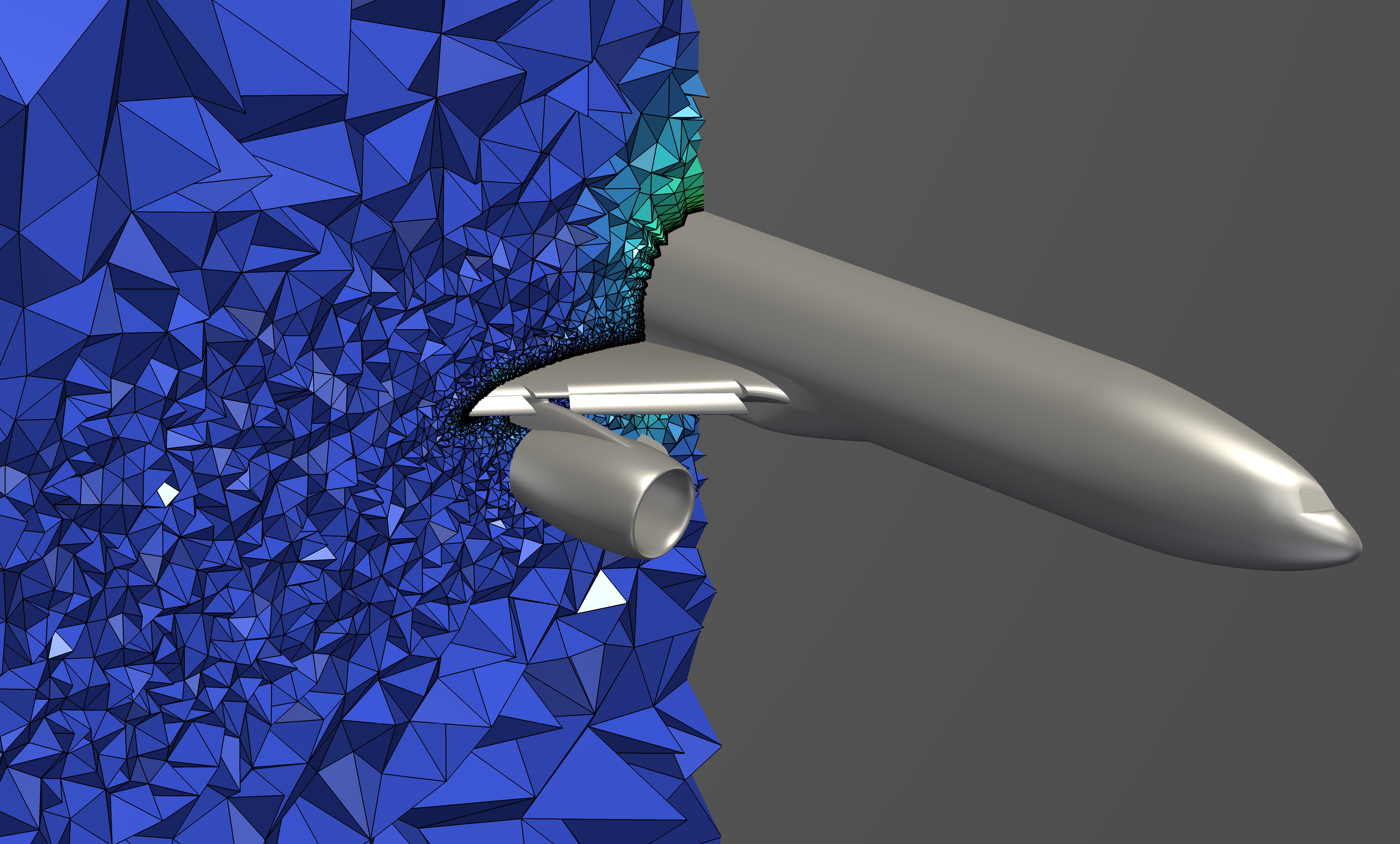}
		\caption{}
		\label{fig:CRM_Y100_P3_Cut}
	\end{subfigure}
	\hfill\hspace{0cm}
	\caption{Volume mesh of the aircraft with $Y+ = 100$ for the polynomial
		degrees:
		(a) $Q = 2$; and
		(b) $Q = 3$.
	}
	\label{fig:CRM_Y100}
\end{figure*}

\begin{figure*}[h!]
	\centering
	\hfill
	\begin{subfigure}[b]{0.90\textwidth}
		\includegraphics[width=\textwidth]{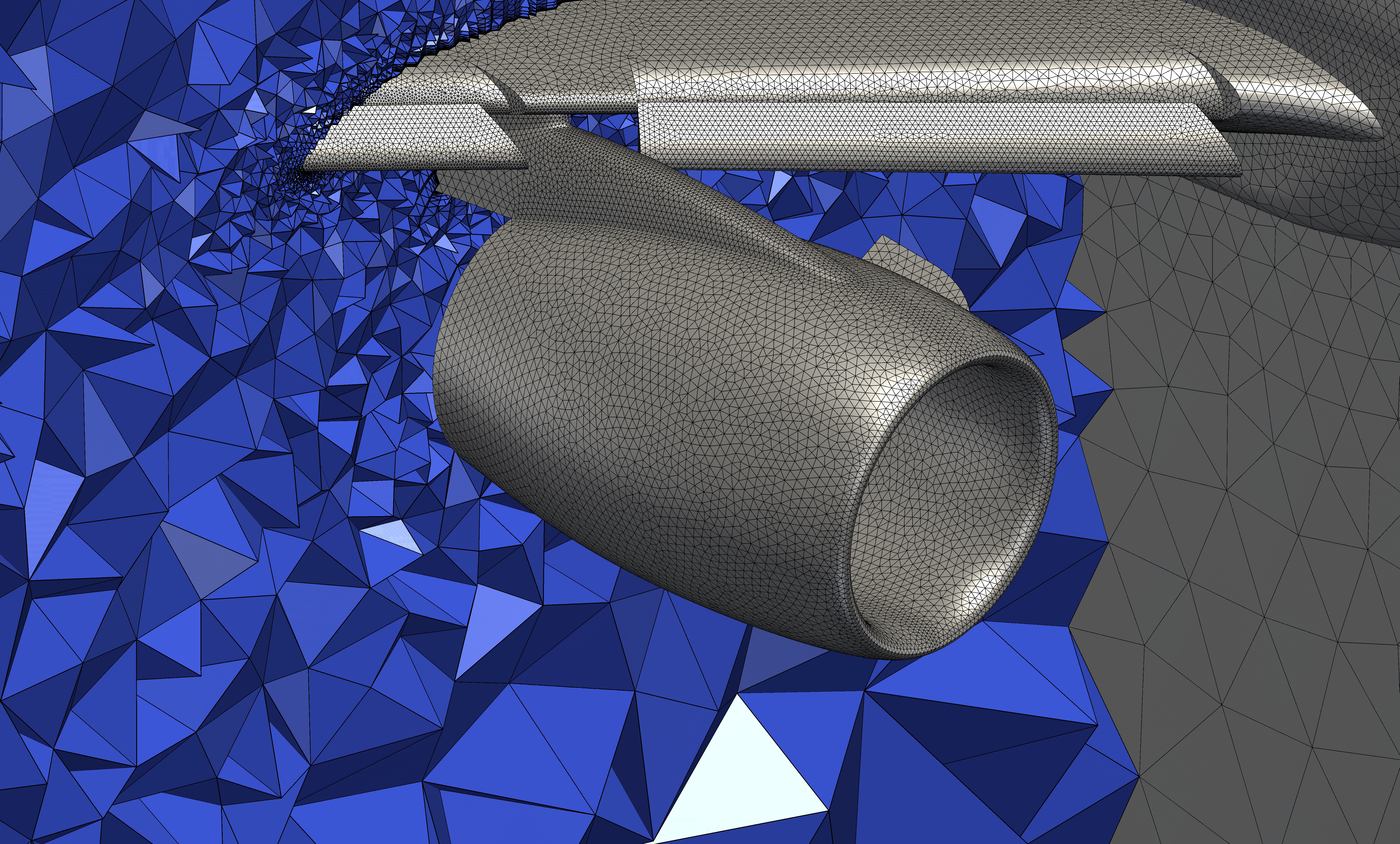}
		\caption{}
		\label{fig:CRM_Y100_P2_Engine}
	\end{subfigure}
	\hfill\hspace{0cm}
	\\
	\hfill
	\begin{subfigure}[b]{0.90\textwidth}
		\includegraphics[width=\textwidth]{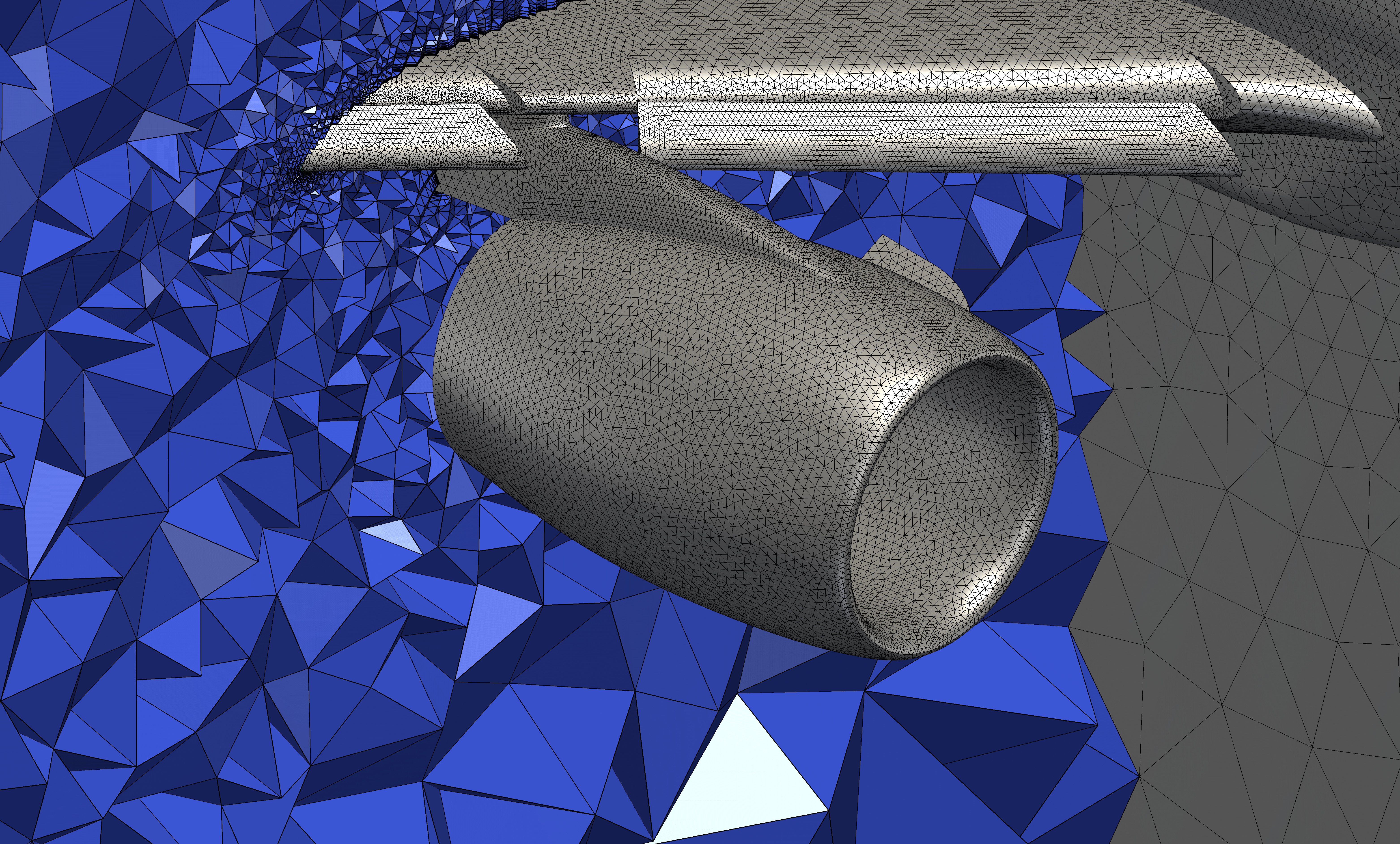}
		\caption{}
		\label{fig:CRM_Y100_P3_Engine}
	\end{subfigure}
	\hfill\hspace{0cm}
	\caption{Detail of the volume mesh around the nacelle with $Y+ = 100$ for the polynomial degrees:
		(a) $Q = 2$; and
		(b) $Q = 3$.
	}
	\label{fig:CRM_Y100_Engine}
\end{figure*}

We show in Figure \ref{fig:CRM_Y100} a slice of the meshes of polynomial degree two and three with $Y+ = 100$. The majority of the elements remain straight-sided, and the curvature of the aircraft smoothly propagates to the interior of the mesh. We show in Figure \ref{fig:CRM_Y100_Engine} a detail of the surface mesh of the nacelle for polynomial degrees two and three. Note that in both cases the surface of the curved mesh is smooth because the highlights of the boundary meshes are also smooth.

\begin{figure*}[h!]
	\centering
	\hfill
	\begin{subfigure}[b]{0.45\textwidth}
		\includegraphics[width=\textwidth]{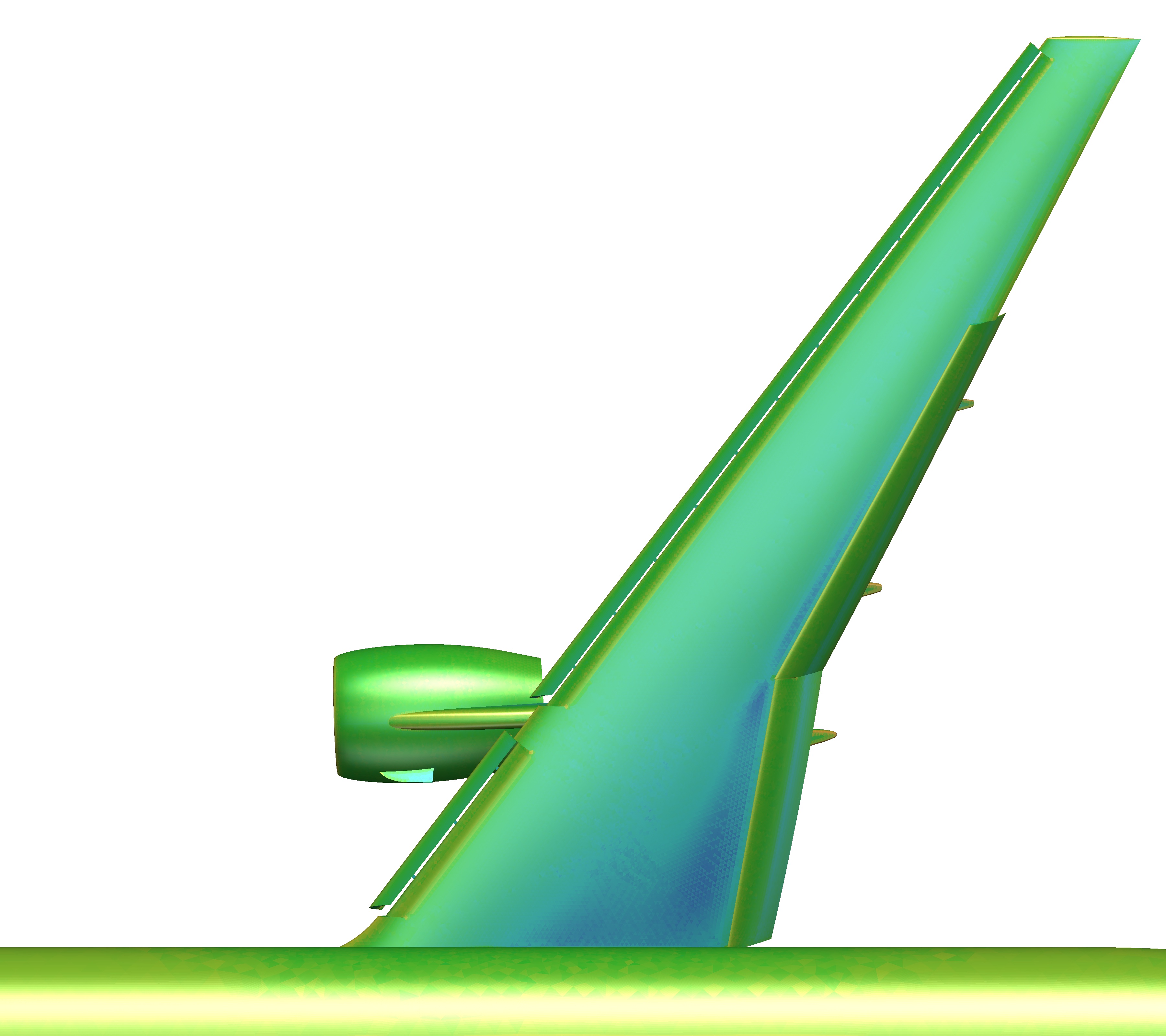}
		\caption{}
		\label{fig:CRM_Y100_P2_View1_SQ}
	\end{subfigure}
	\hfill
	\begin{subfigure}[b]{0.45\textwidth}
		\includegraphics[width=\textwidth]{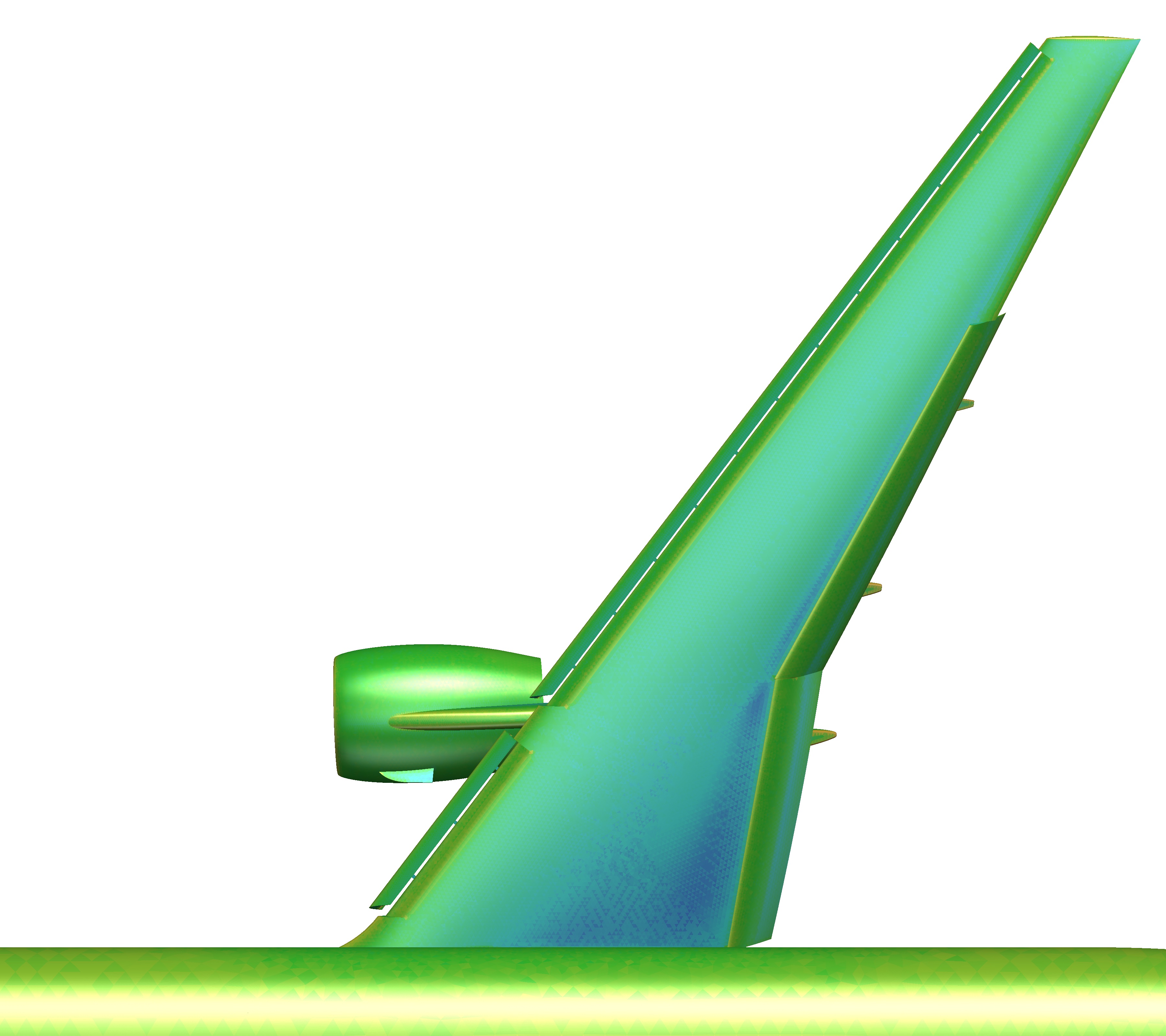}
		\caption{}
		\label{fig:CRM_Y100_P3_View1_SQ}
	\end{subfigure}
	\hfill\hspace{0cm}
	\\
	\hfill
	\begin{subfigure}[b]{0.45\textwidth}
		\includegraphics[width=\textwidth]{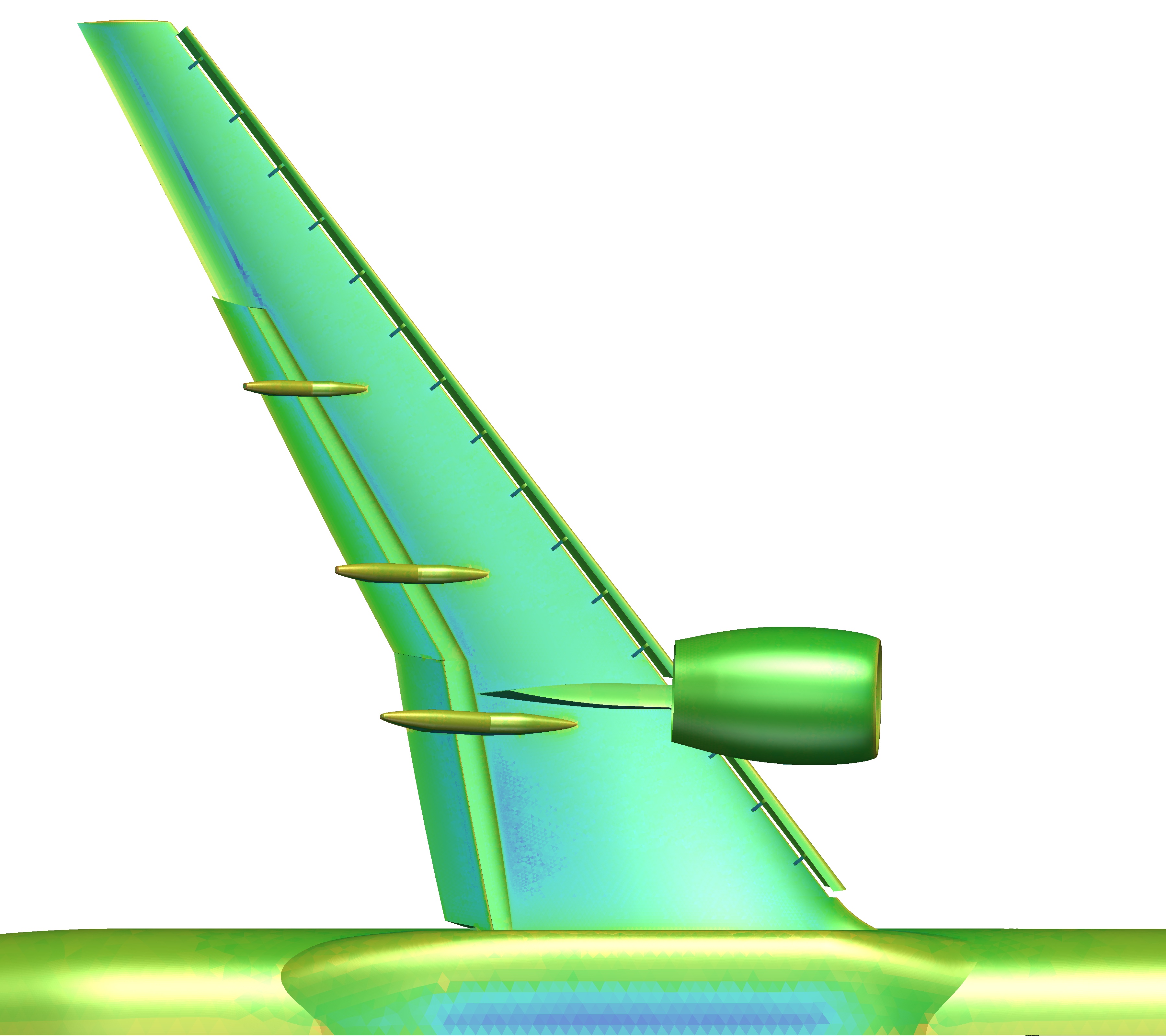}
		\caption{}
		\label{fig:CRM_Y100_P2_View4_SQ}
	\end{subfigure}
	\hfill
	\begin{subfigure}[b]{0.45\textwidth}
		\includegraphics[width=\textwidth]{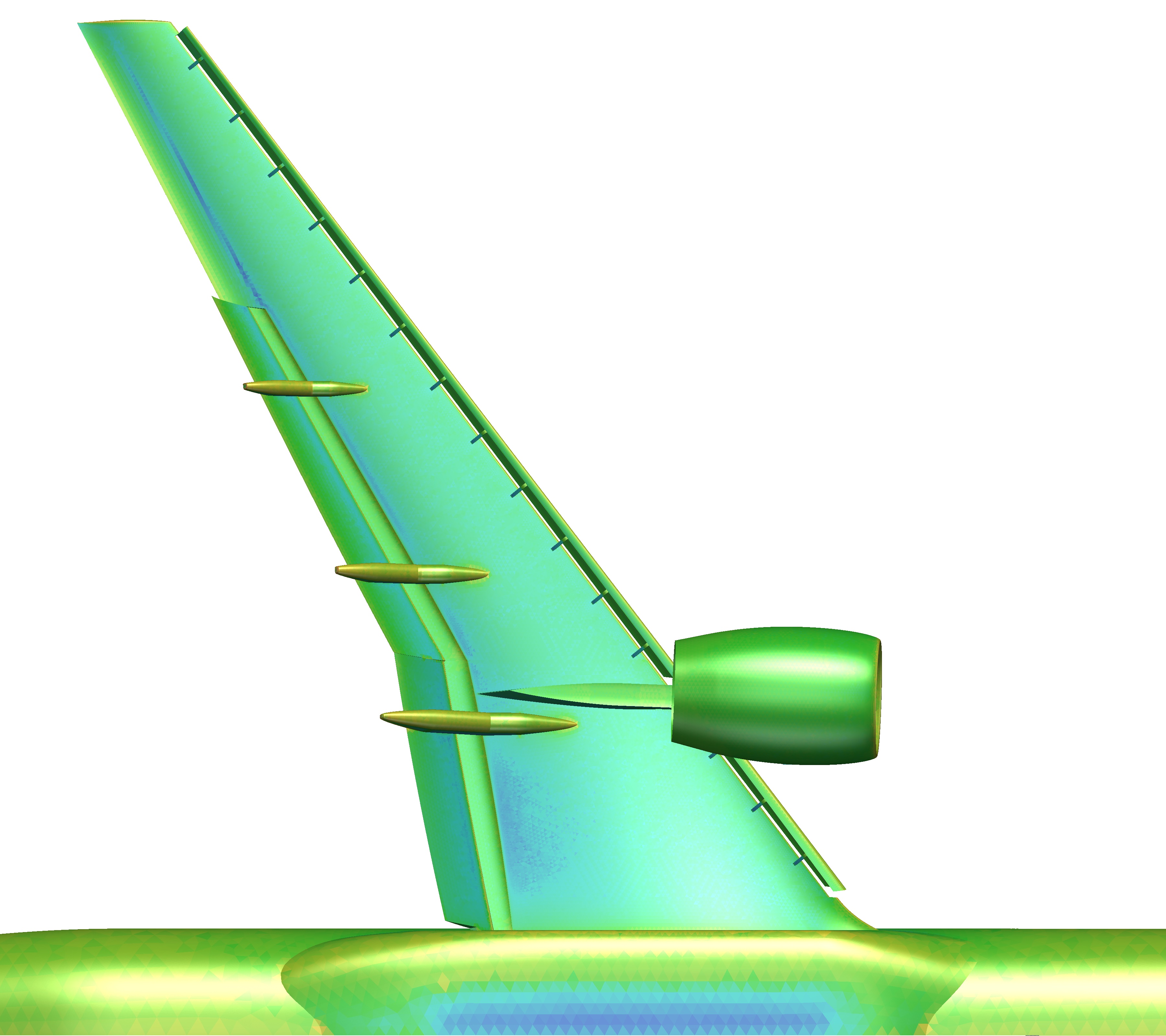}
		\caption{}
		\label{fig:CRM_Y100_P3_View4_SQ}
	\end{subfigure}
	\hfill\hspace{0cm}
	\\
	\hfill
	\begin{subfigure}[b]{0.75\textwidth}
		\includegraphics[width=\textwidth]{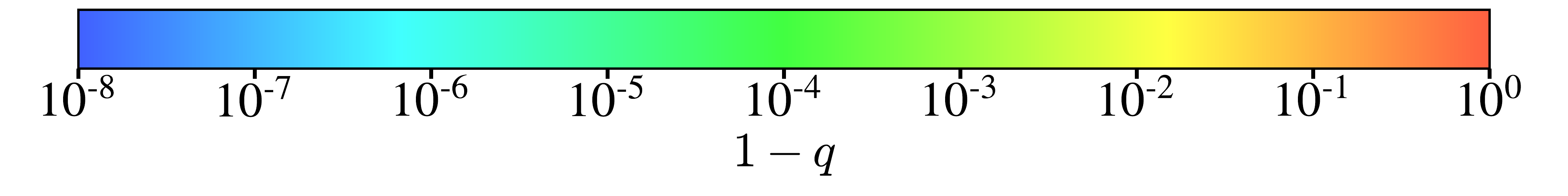}
	\end{subfigure}
	\hfill\hspace{0cm}
	\caption{Shape quality measure of the elements adjacent to the boundary.
		In rows, different views:
		(a) and (b) top view; and
		(c) and (d) bottom view.
		In columns, different polynomial degrees:
		(a) and (c) $Q=2$; and
		(b) and (d) $Q=3$.
	}
	\label{fig:CRM_Y100_Views1-4}
\end{figure*}

\begin{figure*}[h!]
	\centering
	\hfill
	\begin{subfigure}[b]{0.475\textwidth}
		\includegraphics[width=\textwidth]{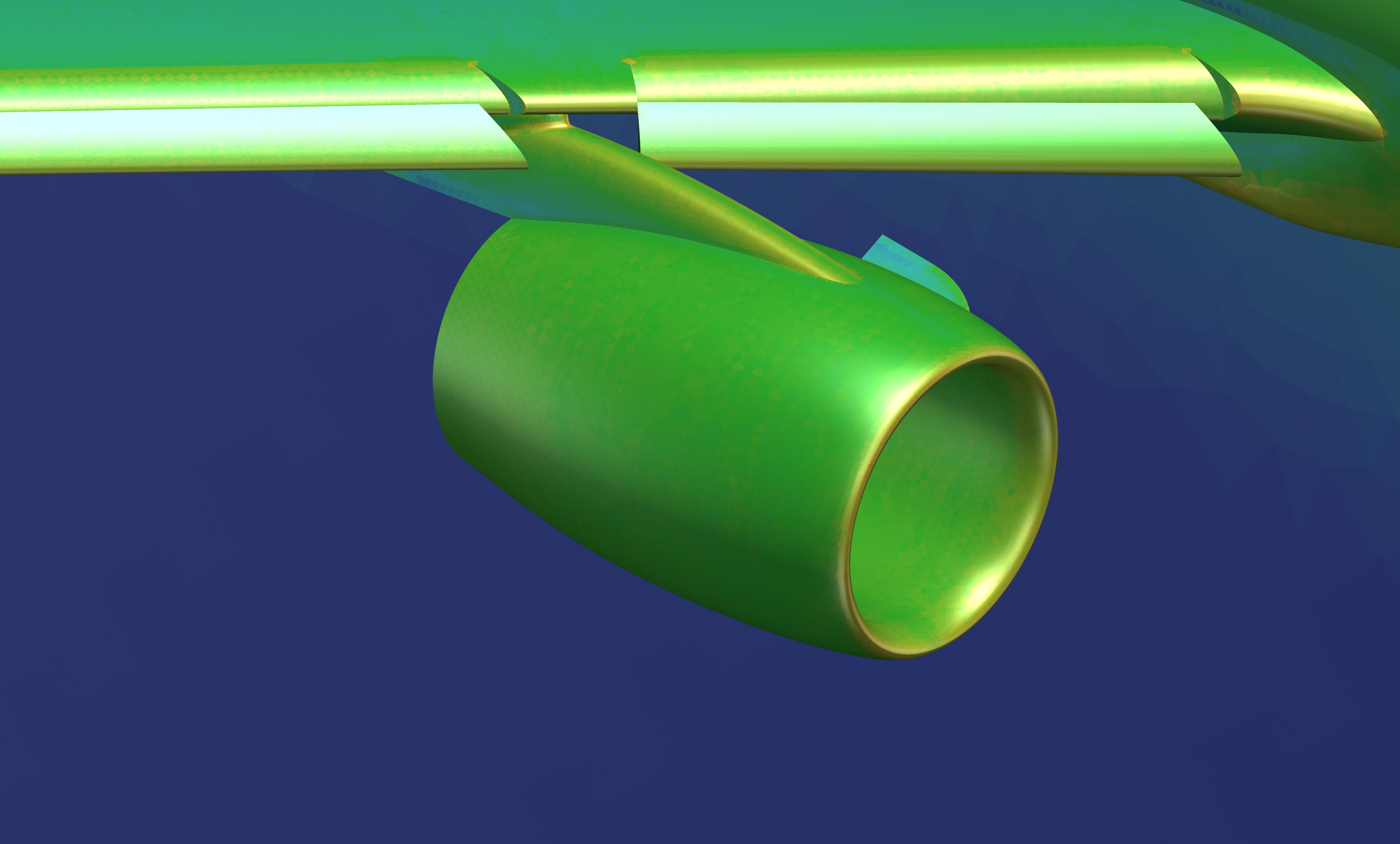}
		\caption{}
		\label{fig:CRM_Y100_P2_ViewEngine_SQ}
	\end{subfigure}
	\hfill
	\begin{subfigure}[b]{0.475\textwidth}
		\includegraphics[width=\textwidth]{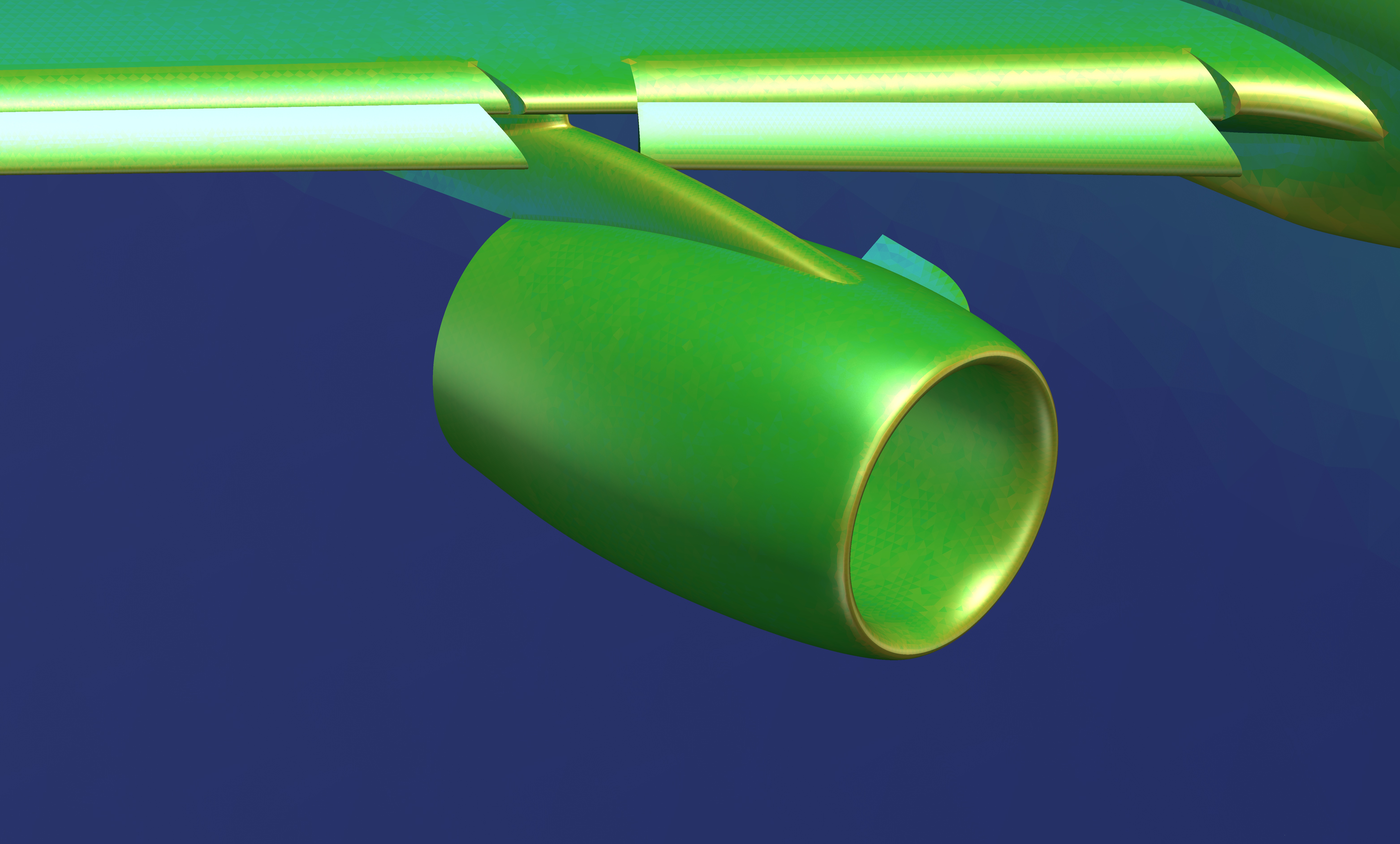}
		\caption{}
		\label{fig:CRM_Y100_P3_ViewEngine_SQ}
	\end{subfigure}
	\hfill\hspace{0cm}
	\\
	\hfill
	\begin{subfigure}[b]{0.75\textwidth}
		\includegraphics[width=\textwidth]{qualityLegend.pdf}
	\end{subfigure}
	\hfill\hspace{0cm}
	\caption{Shape quality measure of the elements adjacent to the nacelle:
		(a) $Q=2$; and
		(b) $Q=3$.
	}
	\label{fig:CRM_Y100_ViewEngine}
\end{figure*}

The lowest quality elements are the ones that are close to the boundary surfaces since they are the most curved ones. The curvature of the elements smoothly decreases the further they are from the boundary. Thus, the quality of the boundary elements is a good indicator of the overall mesh quality. We show in Figure \ref{fig:CRM_Y100_Views1-4} the quality of the boundary elements for two different views of the aircraft with polynomial degrees two and three. Similarly, we show the quality of the boundary elements of the nacelle in Figure \ref{fig:CRM_Y100_ViewEngine}. The quality of the boundary elements is similar in both polynomial degrees. Moreover, the plot shows that as the elements become more curved, their quality decreases. This is expected since the elements have to accommodate the curvature of the boundary.

\begin{figure*}[h!]
	\centering
	\hfill
	\begin{subfigure}[b]{0.50\textwidth}
		\includegraphics[width=\textwidth]{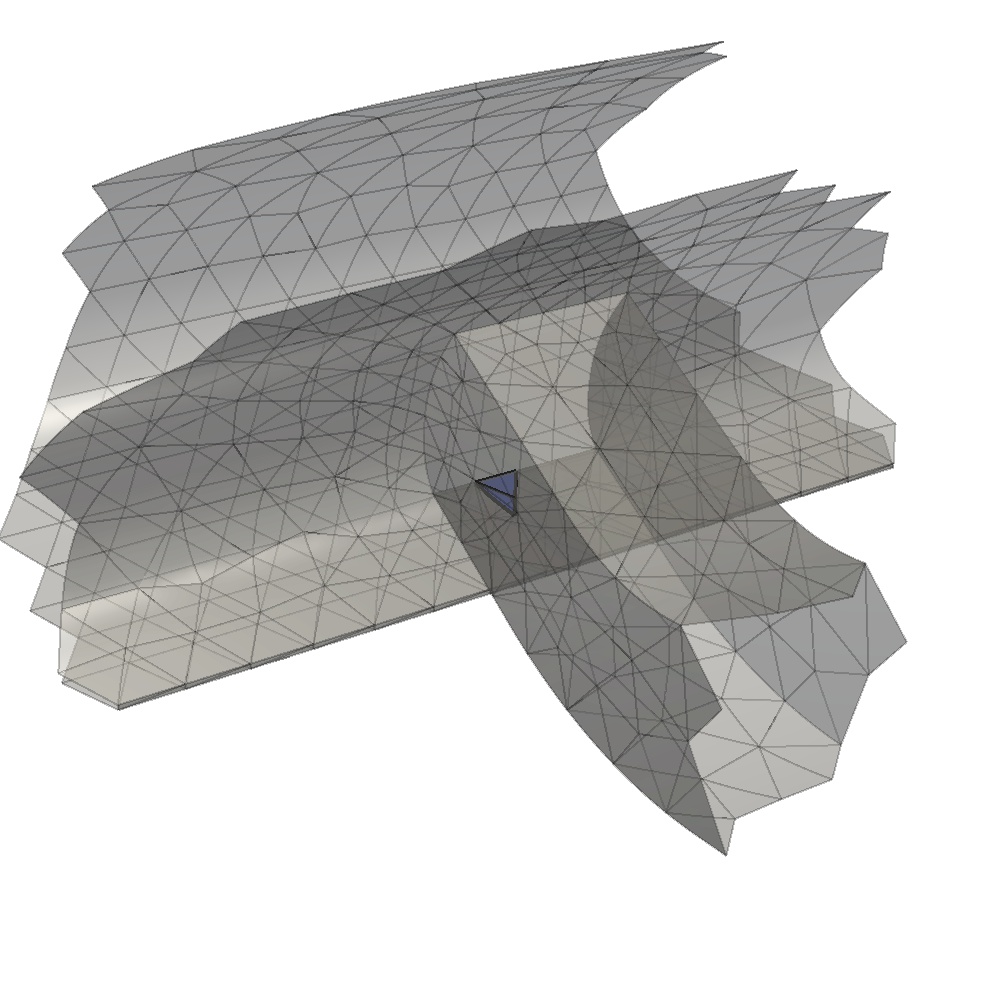}
		\caption{}
		\label{fig:crm_badElement}
	\end{subfigure}
	\hfill
	\begin{subfigure}[b]{0.20\textwidth}
		\includegraphics[width=\textwidth]{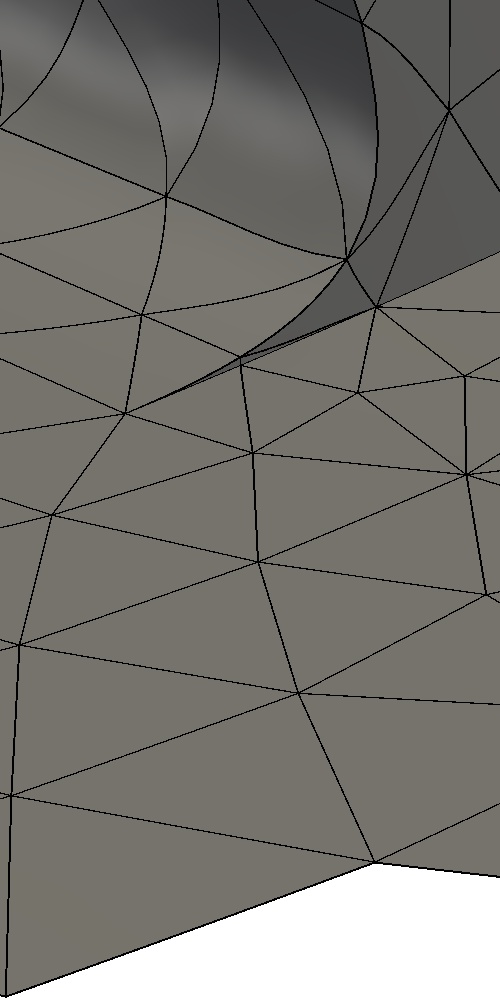}
		\caption{}
		\label{fig:crm_tangentElement}
	\end{subfigure}
	\hfill\hspace{0cm}
	\caption{Geometry configurations that lead to low-quality elements:
		(a) thin region in the wing; and
		(b) tangent curves in the wing.
	}
	\label{fig:crm_badConfigurations}
\end{figure*}

During the visual exploration, we have located the lowest quality element, see Figure \ref{fig:crm_badElement}. This element is located in a thin area of the domain with two surfaces that define a small angle. As a consequence, the feasibility region of the element is small. The geometry configuration makes it difficult to generate a high-quality mesh around this area. The other configuration of the domain that leads to low quality elements is shown in Figure \ref{fig:crm_tangentElement}, in which there are two curves that are tangent. Therefore, the curved triangle features a null Jacobian at the tangent curves, especially when dealing with small element sizes or high polynomial degrees. In this case, the linear and non-linear problems have high condition numbers and become difficult to solve. To avoid this issue, we keep these triangles \emph{frozen} in the initial straight-edged configuration. In this manner, we can generate a curved mesh at the cost of less geometric accuracy in these areas. Another solution could be to modify the geometric model in order to avoid these curve tangencies. Nevertheless, this is a more difficult solution that involves meshing a different model than the rest of the community.

\subsection{Quality metrics of the curved meshes}

\begin{figure*}[h!]
	\centering
	\hfill
	\begin{subfigure}[b]{0.32\textwidth}
		\includegraphics[width=\textwidth]{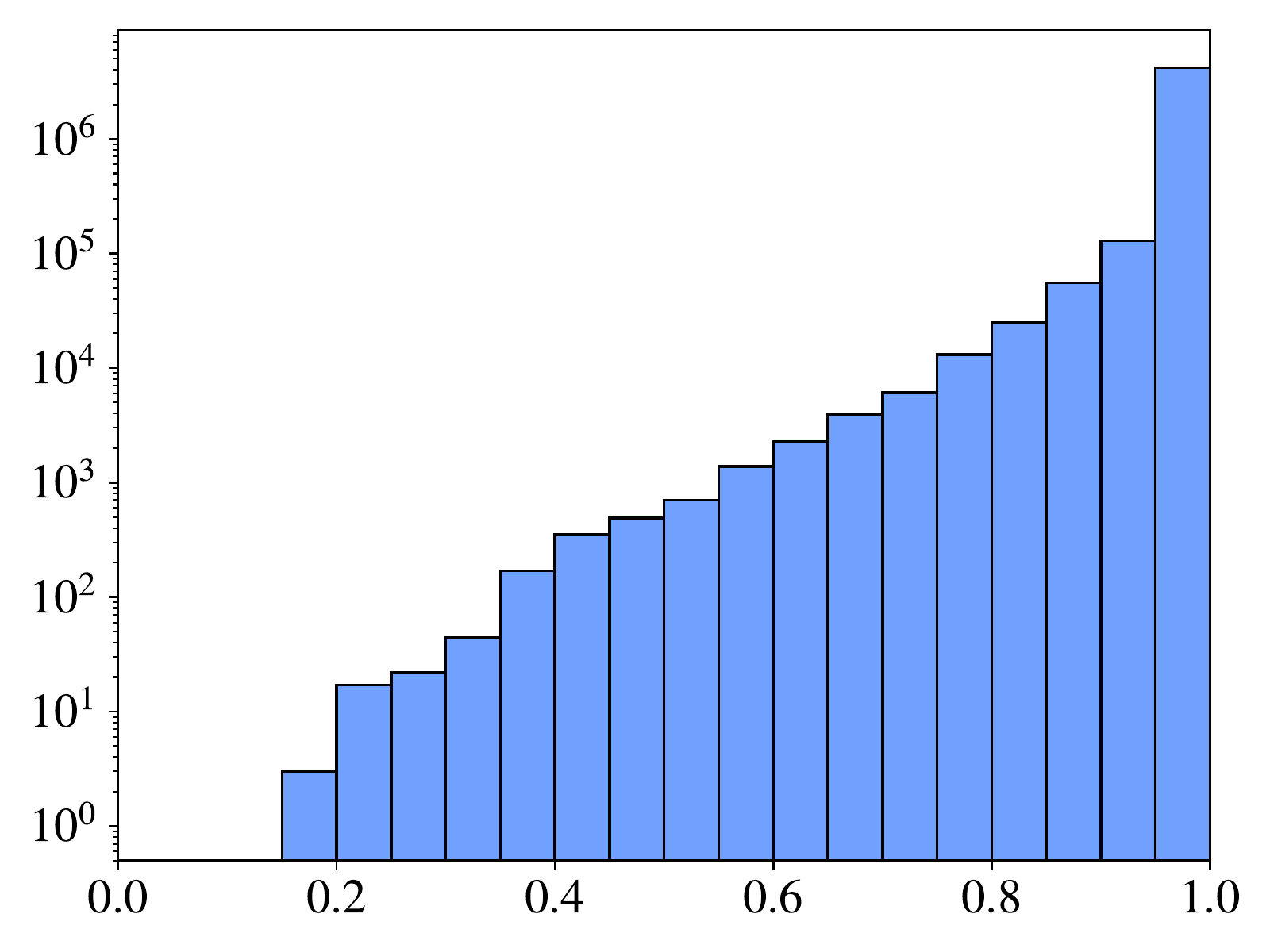}
		\caption{}
		\label{fig:crm_hl_BSC_Coarse_Y800_SJ_P2}
	\end{subfigure}
	\hfill
	\begin{subfigure}[b]{0.32\textwidth}
		\includegraphics[width=\textwidth]{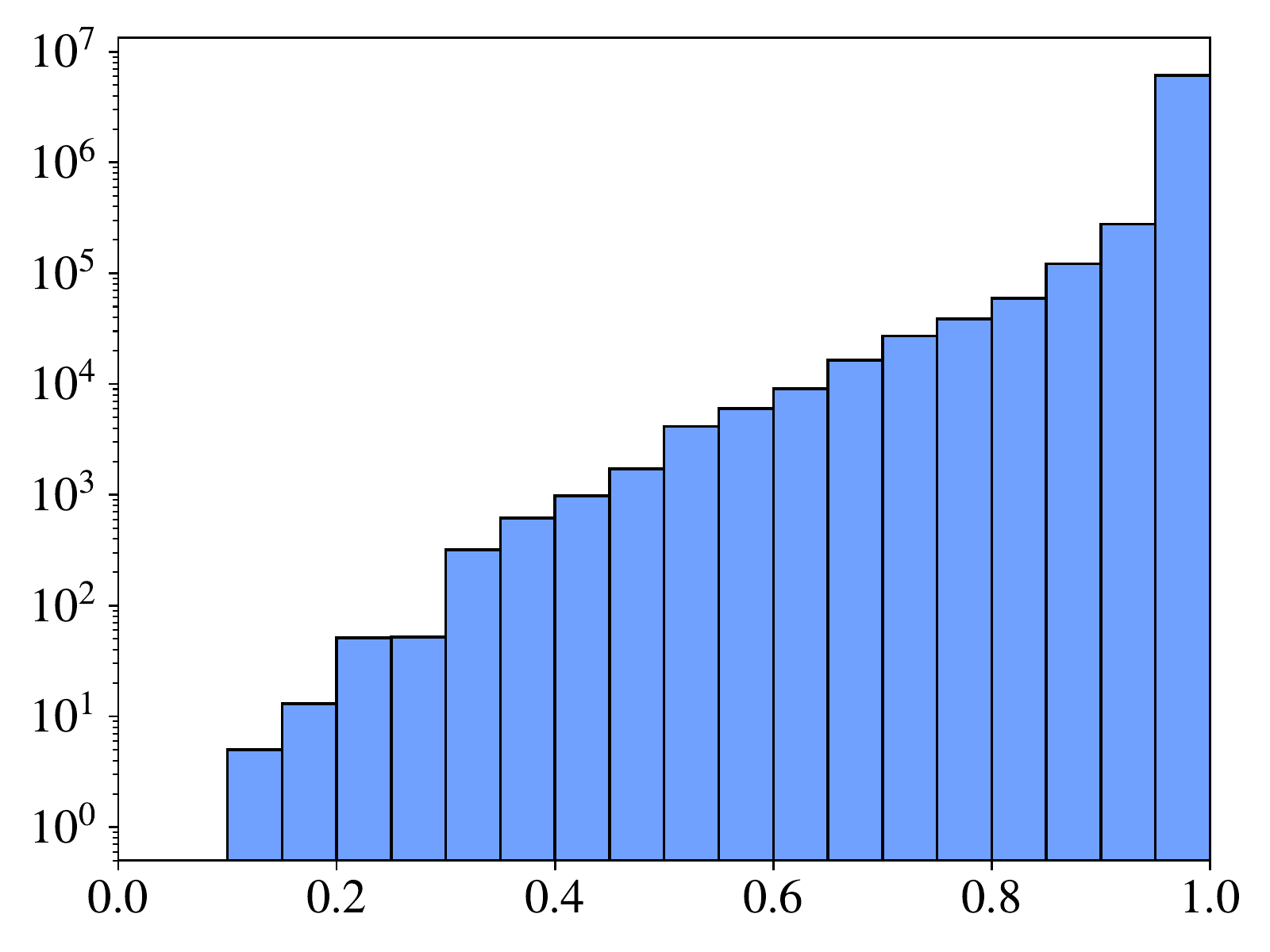}
		\caption{}
		\label{fig:crm_hl_BSC_Coarse_Y200_SJ_P2}
	\end{subfigure}
	\hfill
	\begin{subfigure}[b]{0.32\textwidth}
		\includegraphics[width=\textwidth]{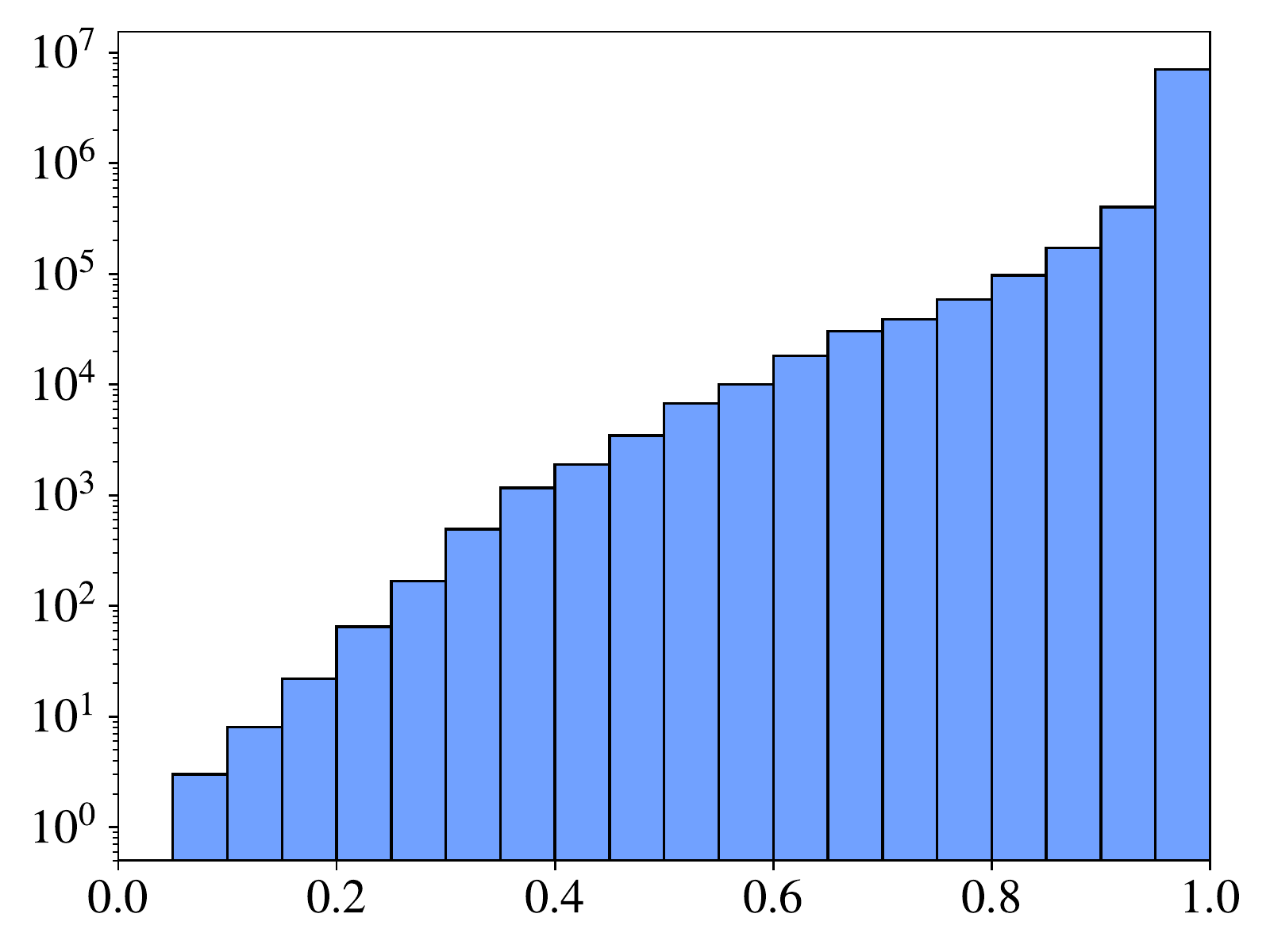}
		\caption{}
		\label{fig:crm_hl_BSC_Coarse_Y100_SJ_P2}
	\end{subfigure}
	\hfill\hspace{0cm}
	\\
	\hfill
	\begin{subfigure}[b]{0.32\textwidth}
		\includegraphics[width=\textwidth]{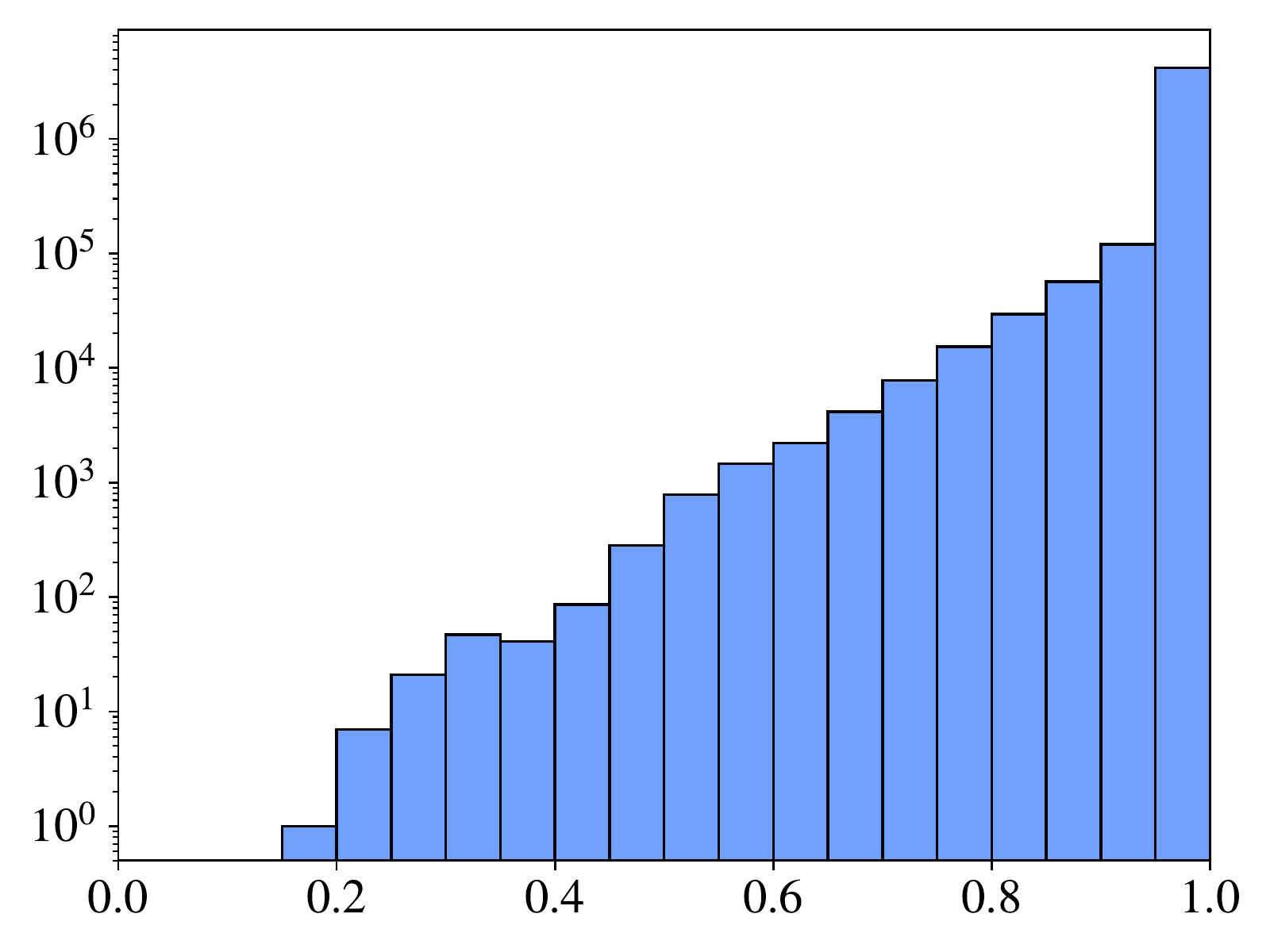}
		\caption{}
		\label{fig:crm_hl_BSC_Coarse_Y800_SJ_P3}
	\end{subfigure}
	\hfill
	\begin{subfigure}[b]{0.32\textwidth}
		\includegraphics[width=\textwidth]{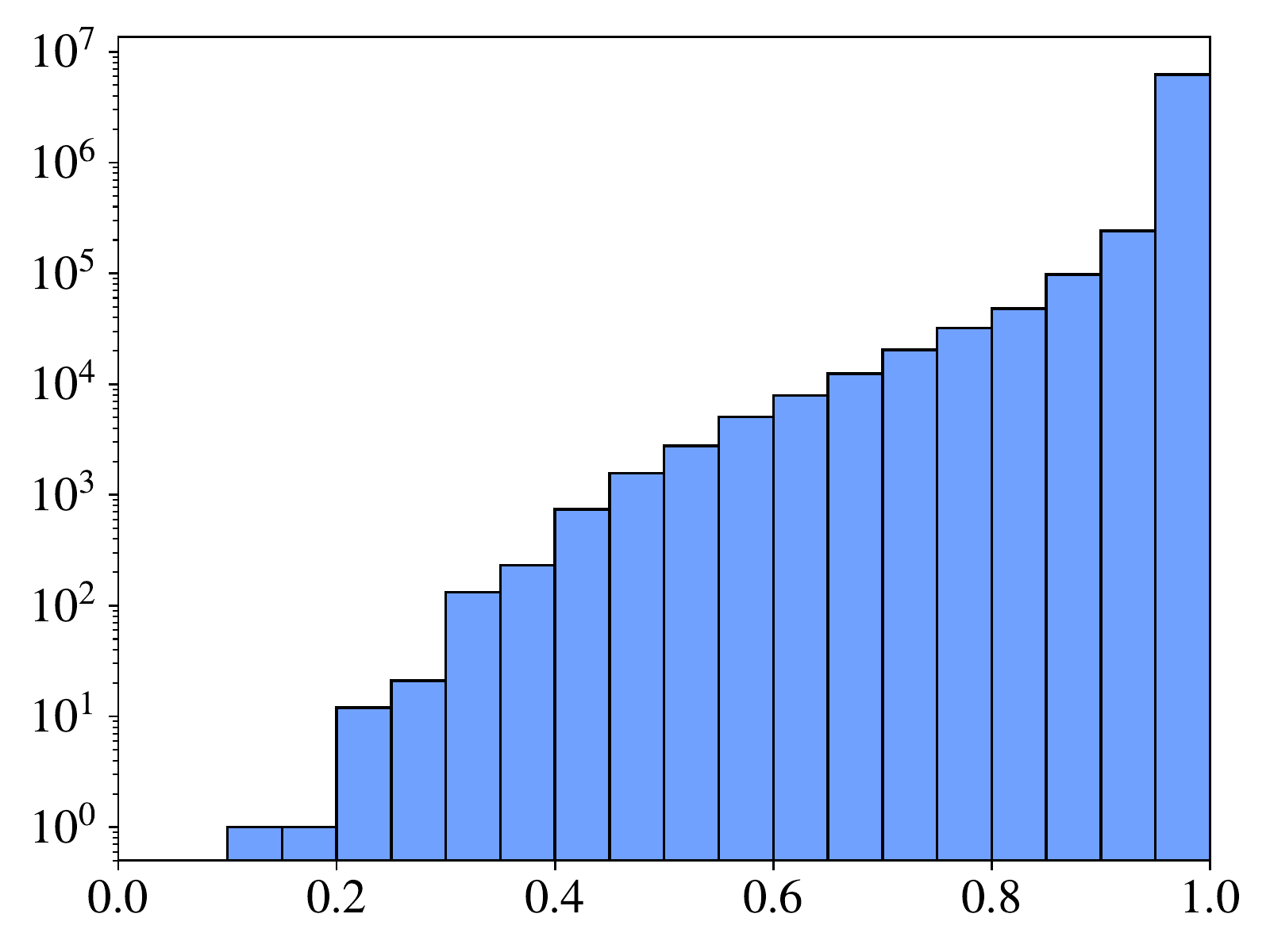}
		\caption{}
		\label{fig:crm_hl_BSC_Coarse_Y200_SJ_P3}
	\end{subfigure}
	\hfill
	\begin{subfigure}[b]{0.32\textwidth}
		\includegraphics[width=\textwidth]{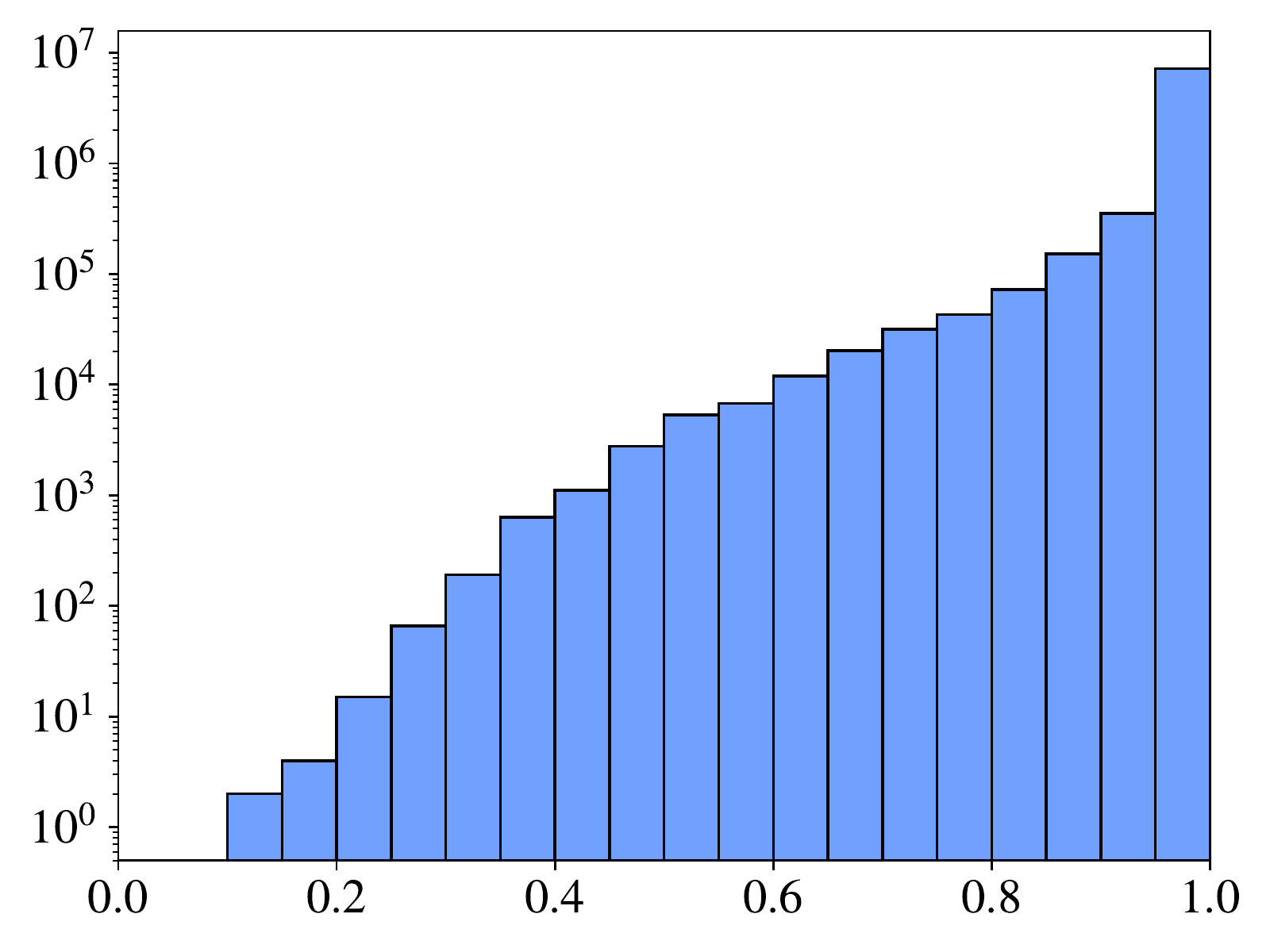}
		\caption{}
		\label{fig:crm_hl_BSC_Coarse_Y100_SJ_P3}
	\end{subfigure}
	\hfill\hspace{0cm}
	\caption{Distribution of elements in logarithmic scale according to their scaled Jacobian.
		In columns, different $Y+$ values:
		(a) and (d) $Y+=800$;
		(b) and (e) $Y+=200$; and
		(c) and (f) $Y+=100$.
		In rows, different polynomial degree:
		(a), (b) and (c) $p=2$; and
		(d), (e) and (f) $p=3$.
	}
	\label{fig:crm_qualityHystogram_SJ}
\end{figure*}

\begin{figure*}[h!]
	\centering
	\hfill
	\begin{subfigure}[b]{0.32\textwidth}
		\includegraphics[width=\textwidth]{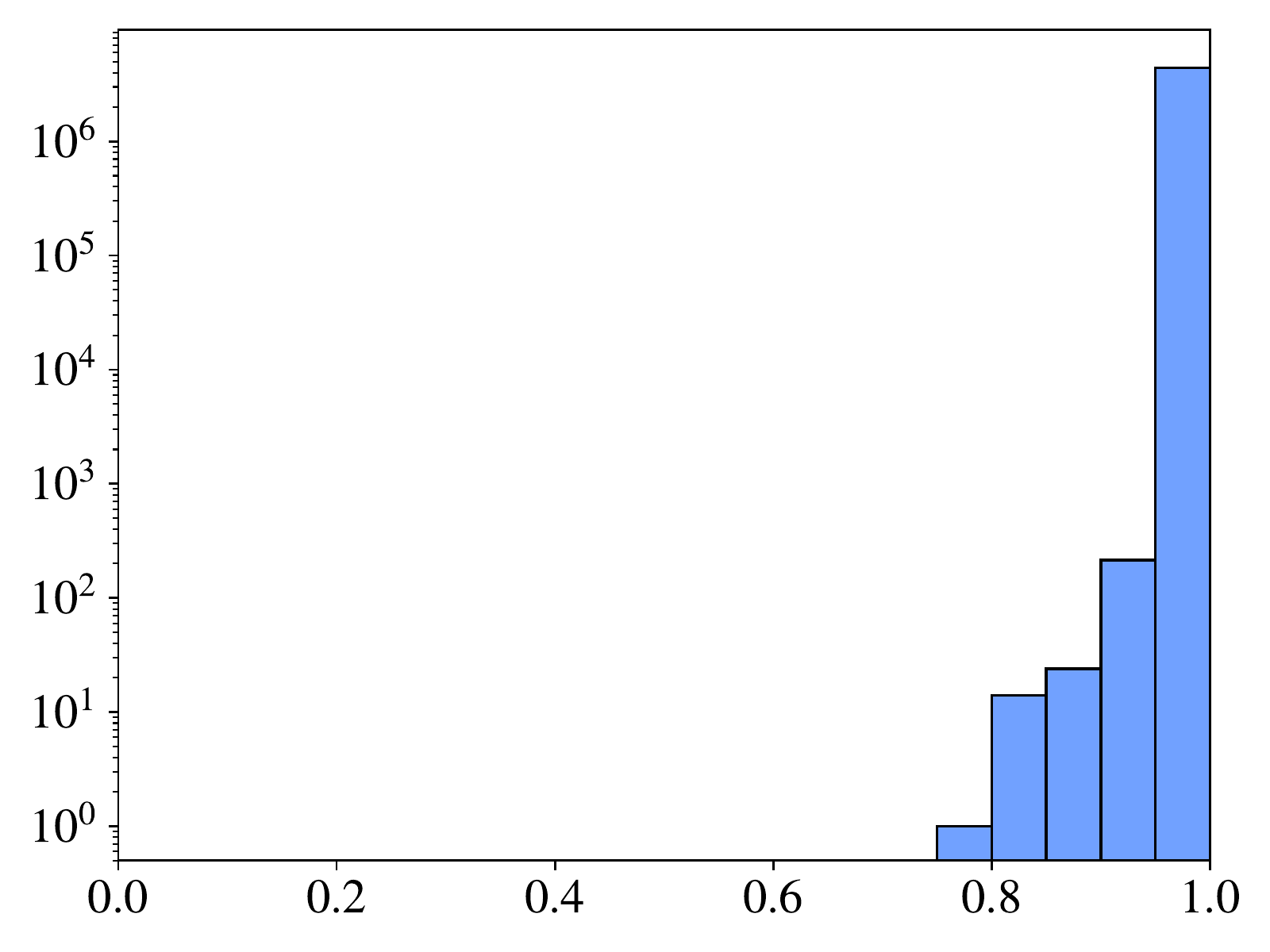}
		\caption{}
		\label{fig:crm_hl_BSC_Coarse_Y800_SQ_P2}
	\end{subfigure}
	\hfill
	\begin{subfigure}[b]{0.32\textwidth}
		\includegraphics[width=\textwidth]{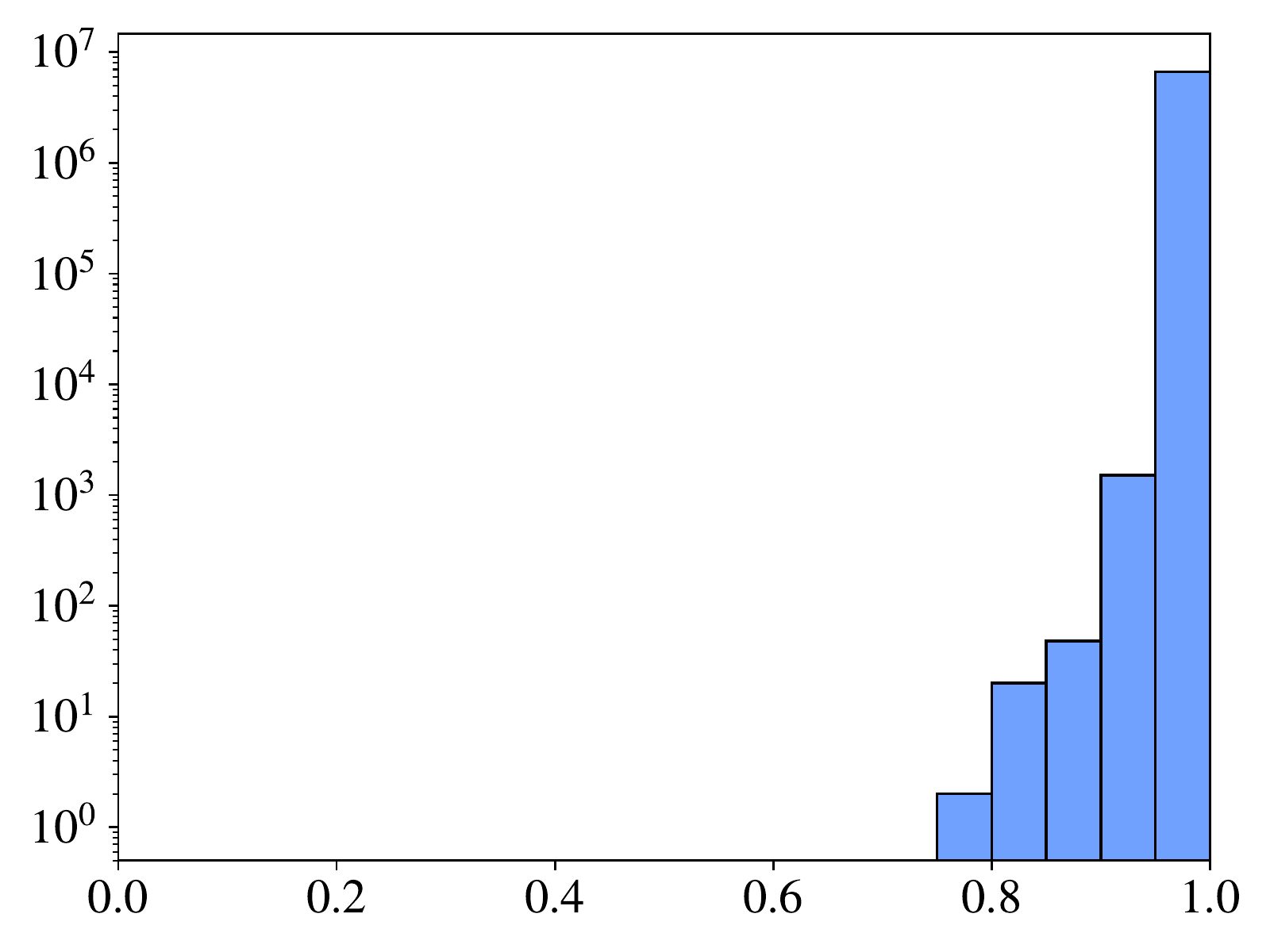}
		\caption{}
		\label{fig:crm_hl_BSC_Coarse_Y200_SQ_P2}
	\end{subfigure}
	\hfill
	\begin{subfigure}[b]{0.32\textwidth}
		\includegraphics[width=\textwidth]{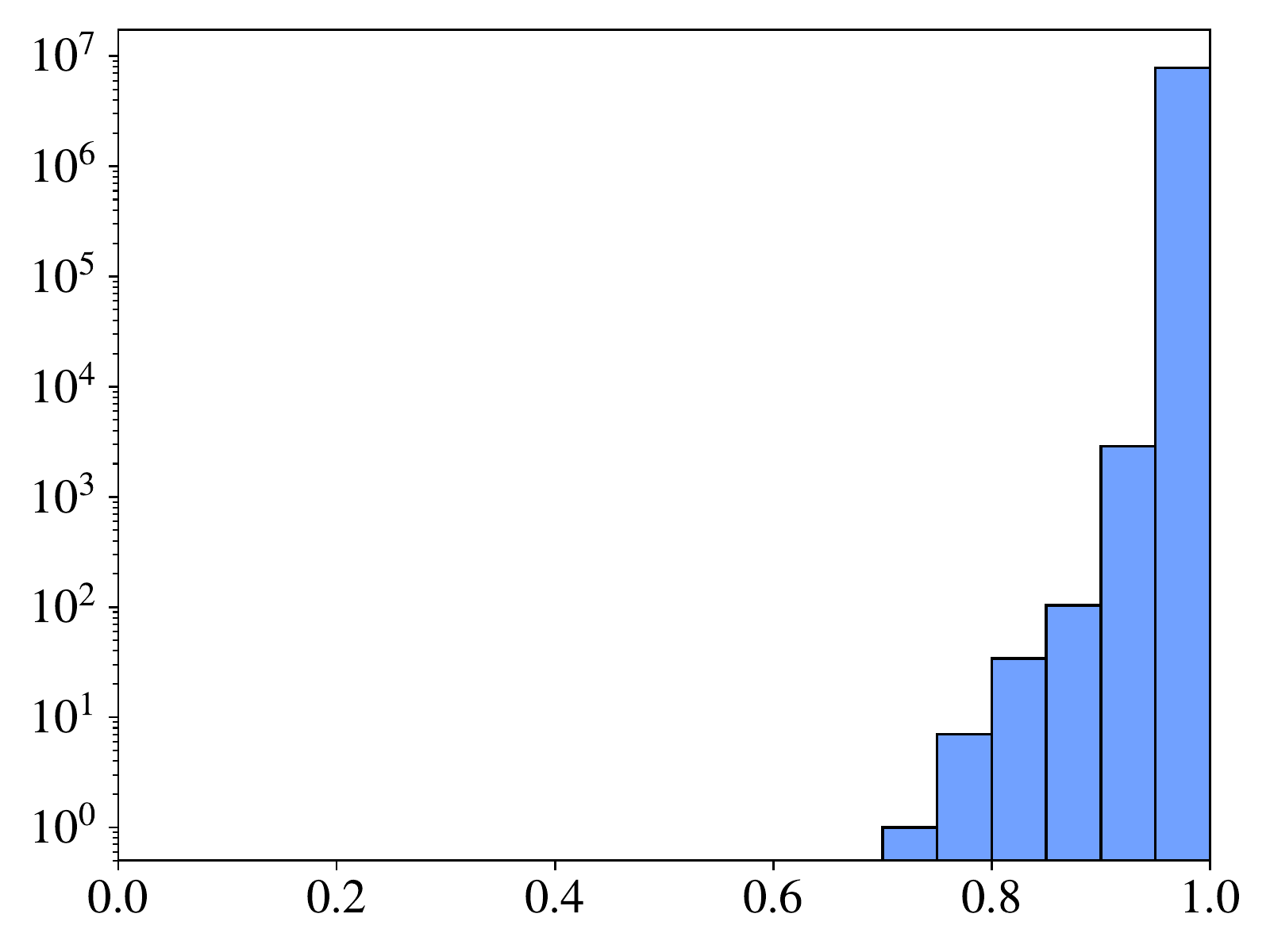}
		\caption{}
		\label{fig:crm_hl_BSC_Coarse_Y100_SQ_P2}
	\end{subfigure}
	\hfill\hspace{0cm}
	\\
	\hfill
	\begin{subfigure}[b]{0.32\textwidth}
		\includegraphics[width=\textwidth]{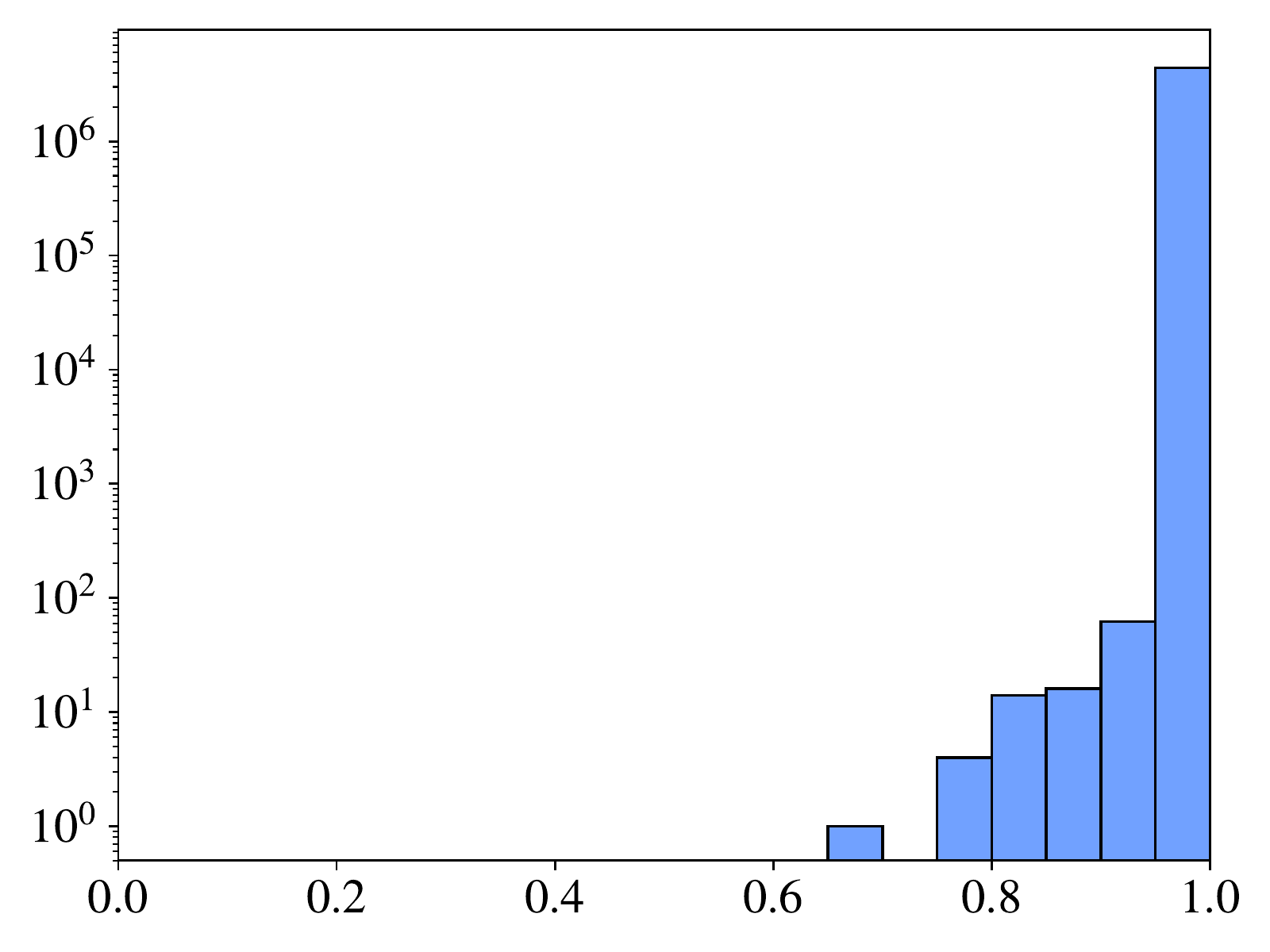}
		\caption{}
		\label{fig:crm_hl_BSC_Coarse_Y800_SQ_P3}
	\end{subfigure}
	\hfill
	\begin{subfigure}[b]{0.32\textwidth}
		\includegraphics[width=\textwidth]{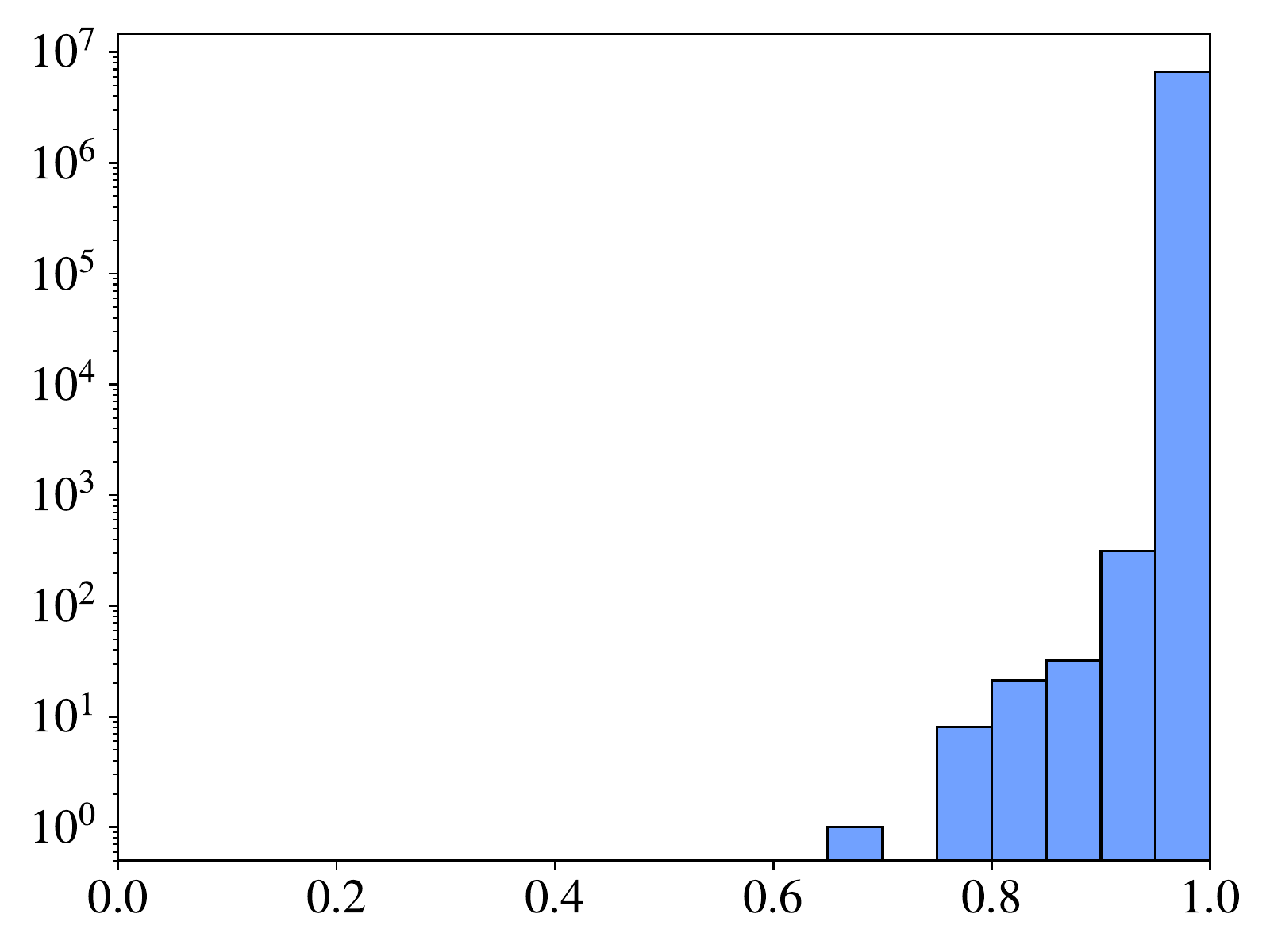}
		\caption{}
		\label{fig:crm_hl_BSC_Coarse_Y200_SQ_P3}
	\end{subfigure}
	\hfill
	\begin{subfigure}[b]{0.32\textwidth}
		\includegraphics[width=\textwidth]{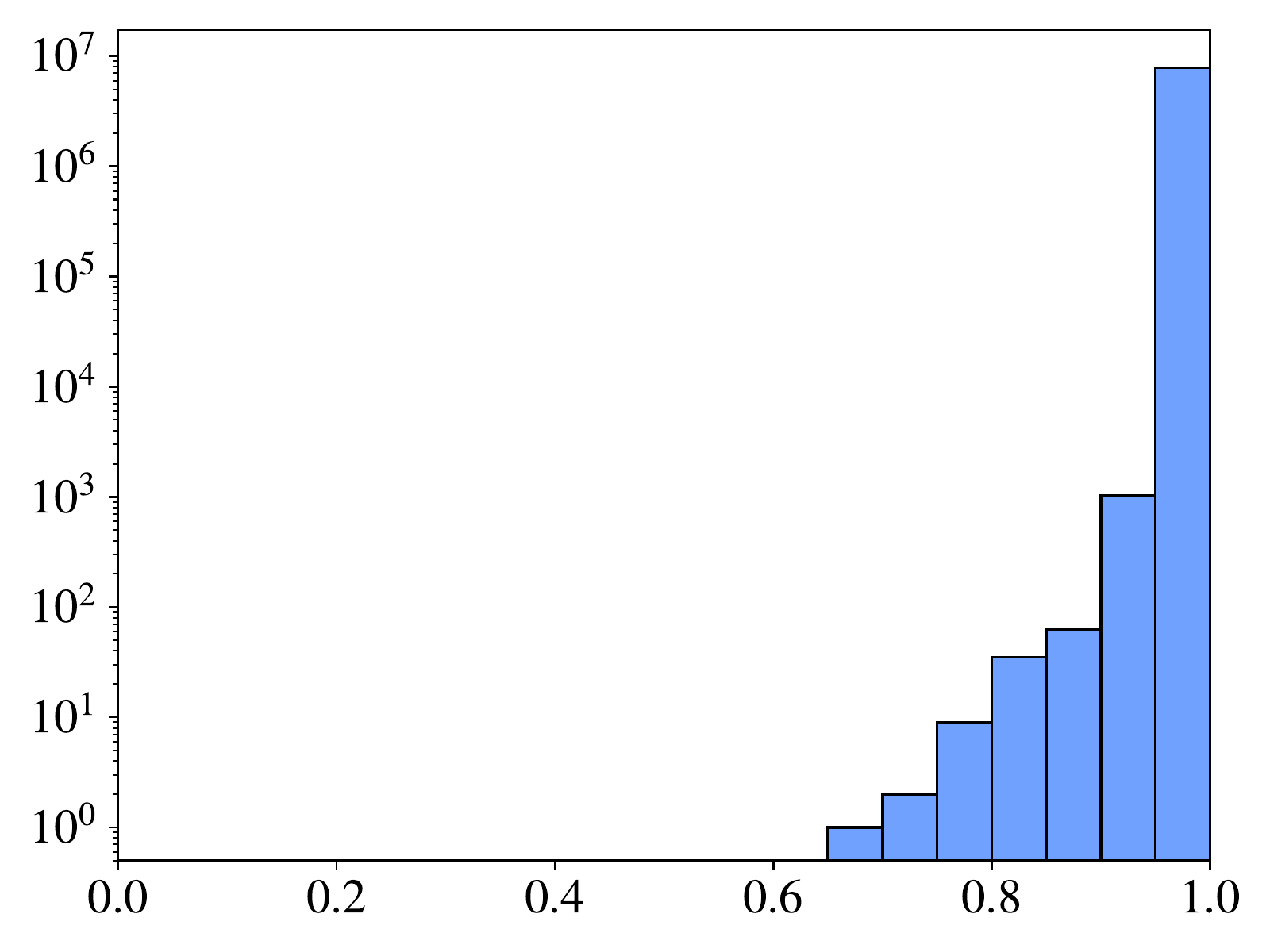}
		\caption{}
		\label{fig:crm_hl_BSC_Coarse_Y100_SQ_P3}
	\end{subfigure}
	\hfill\hspace{0cm}
	\caption{Distribution of elements in logarithmic scale according to their relative shape quality.
		In columns, different $Y+$ values:
		(a) and (d) $Y+=800$;
		(b) and (e) $Y+=200$; and
		(c) and (f) $Y+=100$.
		In rows, different polynomial degree:
		(a), (b) and (c) $p=2$; and
		(d), (e) and (f) $p=3$.
	}
	\label{fig:crm_qualityHystogram_SQ}
\end{figure*}

We analyze the curved meshes in terms of the relative shape quality and the geometric accuracy. We present in Figures \ref{fig:crm_qualityHystogram_SJ} and \ref{fig:crm_qualityHystogram_SQ} the distribution of the elements in logarithmic scale according to the scaled Jacobian and relative shape quality, respectively. In rows, we show polynomial degrees two and three, and in columns we show the $Y+$ values of 800, 200 and 100. The distribution of elements is similar in all the different $Y+$ values and polynomial degrees in which the majority of the elements are in the highest quality bin. Note that all the elements of the mesh are valid since all the element qualities are positive.

\begin{figure*}[h!]
	\centering
	\hfill
	\includegraphics[width=0.45\textwidth]{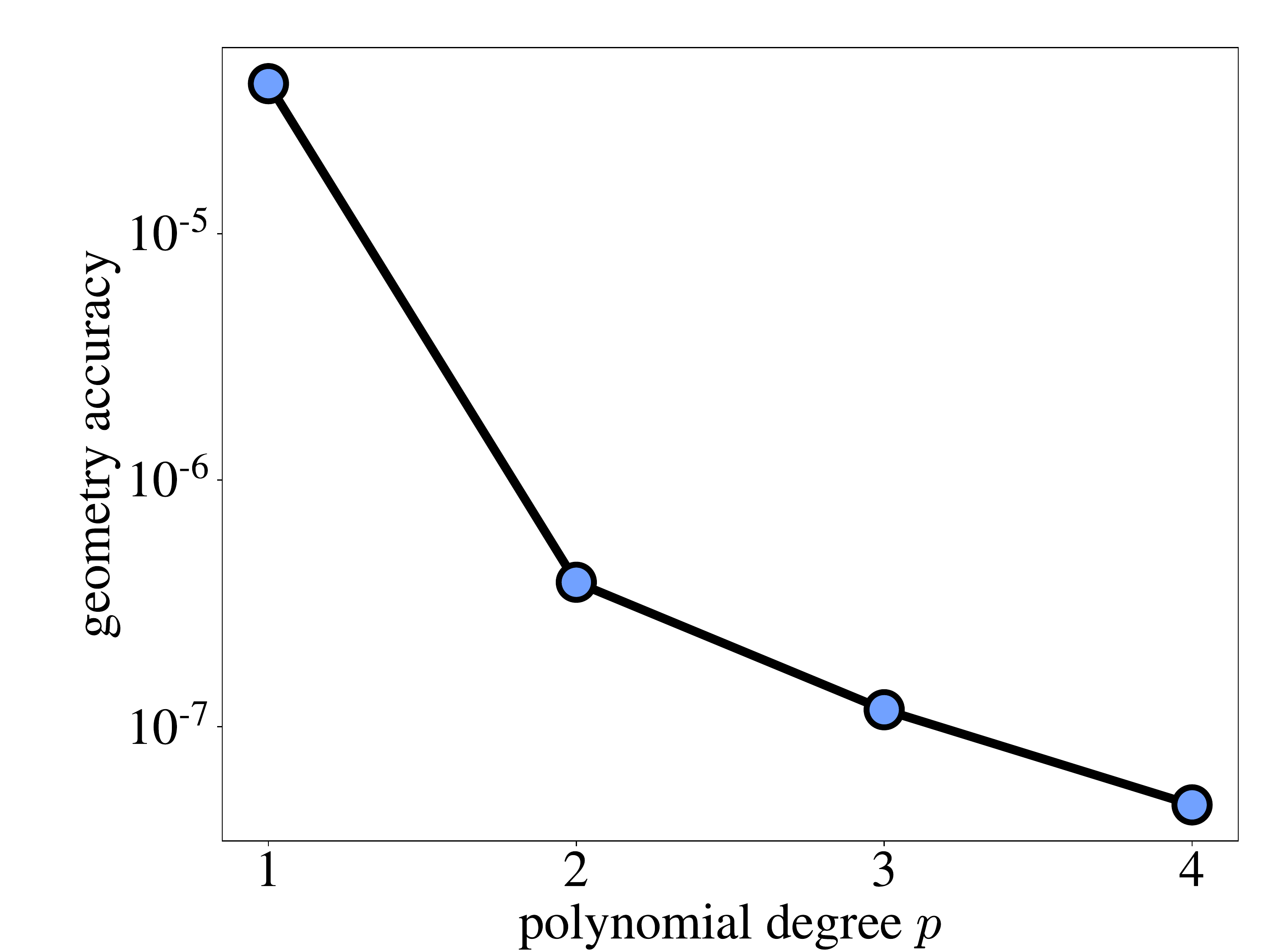}
	\hfill\hspace{0cm}
	\caption{Polynomial degree against the geometric error of the meshes.}
	\label{fig:crm_geometricAccuracy}
\end{figure*}

\subsection{Geometric accuracy}

We show the different geometric accuracy measures for all the generated mesh in Table \ref{tab:distances}. All the measures behave similarly, in the sense that higher polynomial degrees lead to more accurate meshes. Moreover, the different $Y+$ values do not affect the geometric accuracy of the mesh. In particular, all the meshes have the same boundary mesh. As a consequence, we conclude that the geometric accuracy is mainly determined by the boundary mesh. Nevertheless, the volume mesh may introduce small perturbations during the curving process that may modify the geometric accuracy. 

As we increase the polynomial degree, the resulting mesh becomes more accurate, see Figure \ref{fig:crm_geometricAccuracy}. The difference of the geometric accuracy between the initial linear mesh and the quadratic mesh is of two orders of magnitude. This is the largest difference between two consecutive polynomial degrees. Note that we do not obtain an exponential convergence of the geometric accuracy since the slope of the curve is not constant. This means that there are regions of the geometry that are not in the convergence area and we need a smaller element size. Nevertheless, even if we do not have an exponential convergence, the geometric error decreases when we increase the polynomial degree.

\begin{table}[t]
\centering
\caption{Geometric accuracy measures for the different meshes generated, where $\ell_c = 2470$ is the aircraft length}
\label{tab:distances}
\begin{tabular}{*{8}{c}}
	\hline
	& \multicolumn{3}{c}{$Q=2$} & & \multicolumn{3}{c}{$Q=3$}\\
	& $Y+=800$ & $Y+=200$ & $Y+=100$ & \hspace{0.33cm} & $Y+=800$ & $Y+=200$ & $Y+=100$ \\
	\hline
	SC      & $9.82 \cdot 10^{-4}$ & $9.84 \cdot 10^{-4}$ & $9.82 \cdot 10^{-4}$ & & $2.90 \cdot 10^{-4}$ & $2.90 \cdot 10^{-4}$ & $2.89 \cdot 10^{-4}$  \\
	SC/$\ell_c$   & $3.97 \cdot 10^{-7}$ & $3.98 \cdot 10^{-7}$ & $3.97 \cdot 10^{-7}$ & & $1.17 \cdot 10^{-7}$ & $1.17 \cdot 10^{-7}$ & $1.17 \cdot 10^{-7}$  \\
	$d_2$      & $5.13 \cdot 10^{-3}$ & $5.14 \cdot 10^{-3}$ & $5.14 \cdot 10^{-3}$ & & $1.77 \cdot 10^{-3}$ & $1.77 \cdot 10^{-3}$ & $1.77 \cdot 10^{-3}$  \\
	$d_2/\ell_c$   & $2.07 \cdot 10^{-6}$ & $2.08 \cdot 10^{-6}$ & $2.08 \cdot 10^{-6}$ & & $7.19 \cdot 10^{-7}$ & $7.19 \cdot 10^{-7}$ & $7.19 \cdot 10^{-7}$  \\
	$d_\infty$     & $2.40 \cdot 10^{-1}$ & $2.52 \cdot 10^{-1}$ & $2.52 \cdot 10^{-1}$ & & $1.29 \cdot 10^{-1}$ & $1.29 \cdot 10^{-1}$ & $1.21 \cdot 10^{-1}$  \\
	$d_\infty/\ell_c$  & $9.74 \cdot 10^{-5}$ & $1.02 \cdot 10^{-4}$ & $1.02 \cdot 10^{-4}$ & & $5.23 \cdot 10^{-5}$ & $5.22 \cdot 10^{-5}$ & $4.91 \cdot 10^{-5}$ \\
	\hline
\end{tabular}
\end{table}

\subsection{Computational time}

To generate the meshes for the high-lift common research model, we have used a distributed parallel environment with 768 processors. The wall-clock time to curve the meshes is of the order of minutes in all the cases, see Table \ref{tab:timings}. This wall-clock time includes the time spent reading the initial linear mesh, the boundary marks and the CAD model, computing the solution of the curving problem, and writing the final curved mesh.

The wall-clock time of curving the quadratic meshes is low compared to the time to curve the cubic meshes. This is important since the generation of a quadratic mesh allows us to detect mesh artifacts that we need to correct. Consequently, we can improve in a fast manner both the definition of the virtual model and the initial linear mesh to obtain better curved meshes.

With our mesh curving methodology, the main bottleneck to obtain a curved high-order mesh is the definition of an appropriate virtual model and the generation of the initial linear mesh. The time spent in these two steps is measured in days, while the time spent in the curving computation is measured in minutes.

\begin{table}[t]
	\caption{Wall-clock time to curve the linear meshes in seconds.}
	\label{tab:timings}
	\centering
	\begin{tabular}{c*{4}{c}}
		       & Isotropic & $Y+=800$ & $Y+=200$ & $Y+=100$ \\
		\hline
		$p=2$  &  237      &   260    &   372    &    726    \\
		$p=3$  &  1336     &   1233   &   1548   &    2894    \\
	\end{tabular}
\end{table}

\section{Answers to the HO-TFG mesh curving questions}
\label{sec:answers}

\subsection{Can 3D curved meshes be generated for the CRM-HL?}

We have shown that we can generate curved meshes for the CRM-HL. Nevertheless, there are several aspects related to the virtual geometry and the linear mesh that we have to take into account.

\begin{figure*}[t!]
	\centering
	\hfill
	\begin{subfigure}[b]{0.27\textwidth}
		\includegraphics[width=\textwidth]{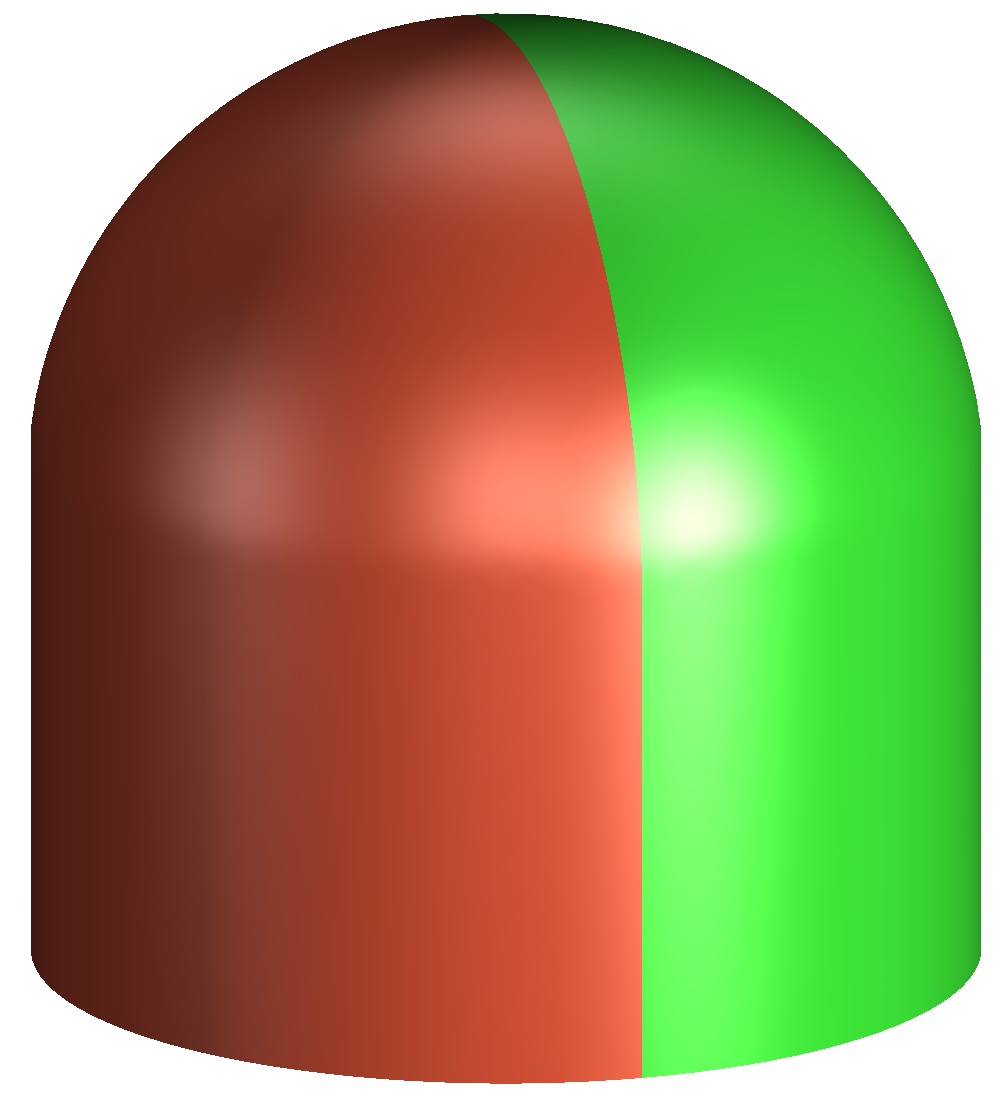}
		\caption{}
		\label{fig:bullet_bad_CAD}
	\end{subfigure}
	\hfill
	\begin{subfigure}[b]{0.27\textwidth}
		\includegraphics[width=\textwidth]{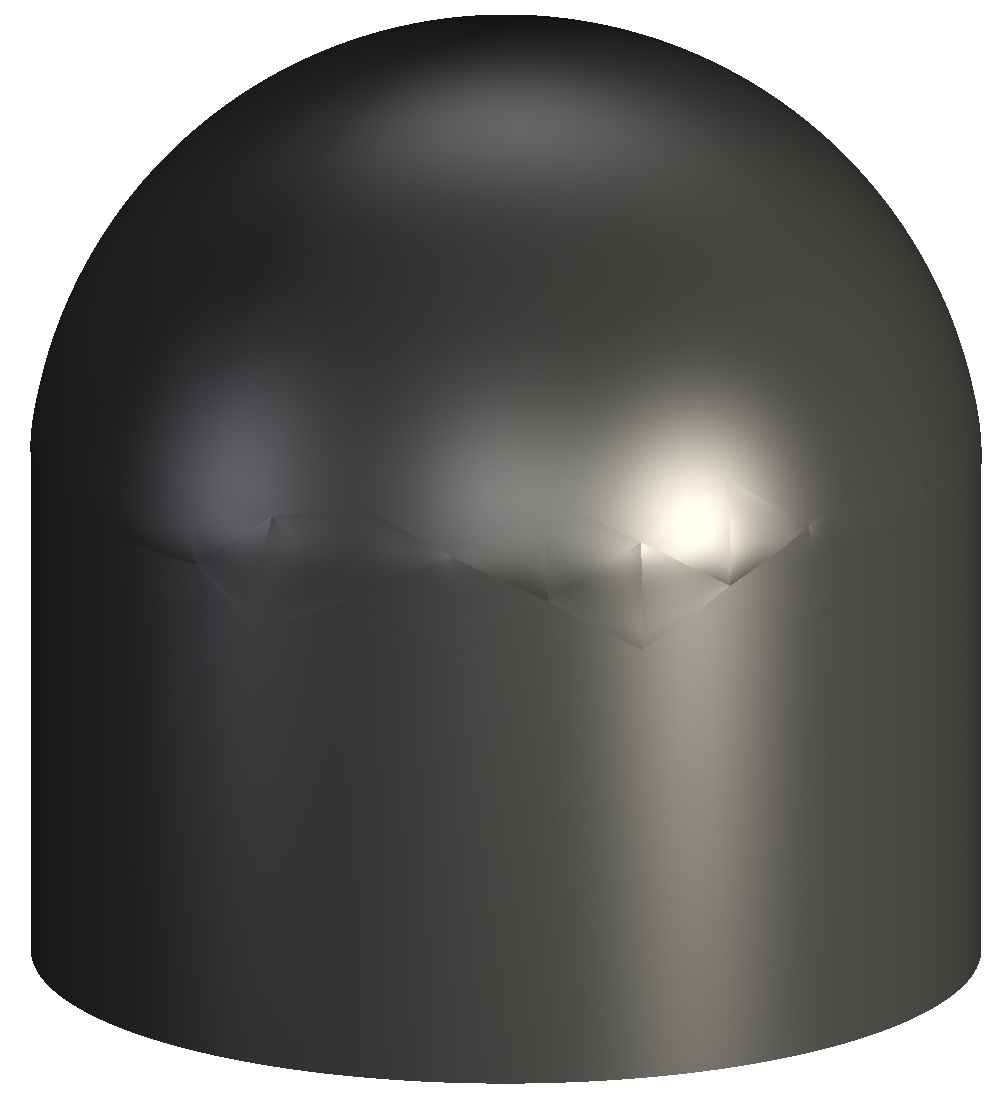}
		\caption{}
		\label{fig:bullet_bad_mesh}
	\end{subfigure}
	\hfill
	\begin{subfigure}[b]{0.27\textwidth}
		\includegraphics[width=\textwidth]{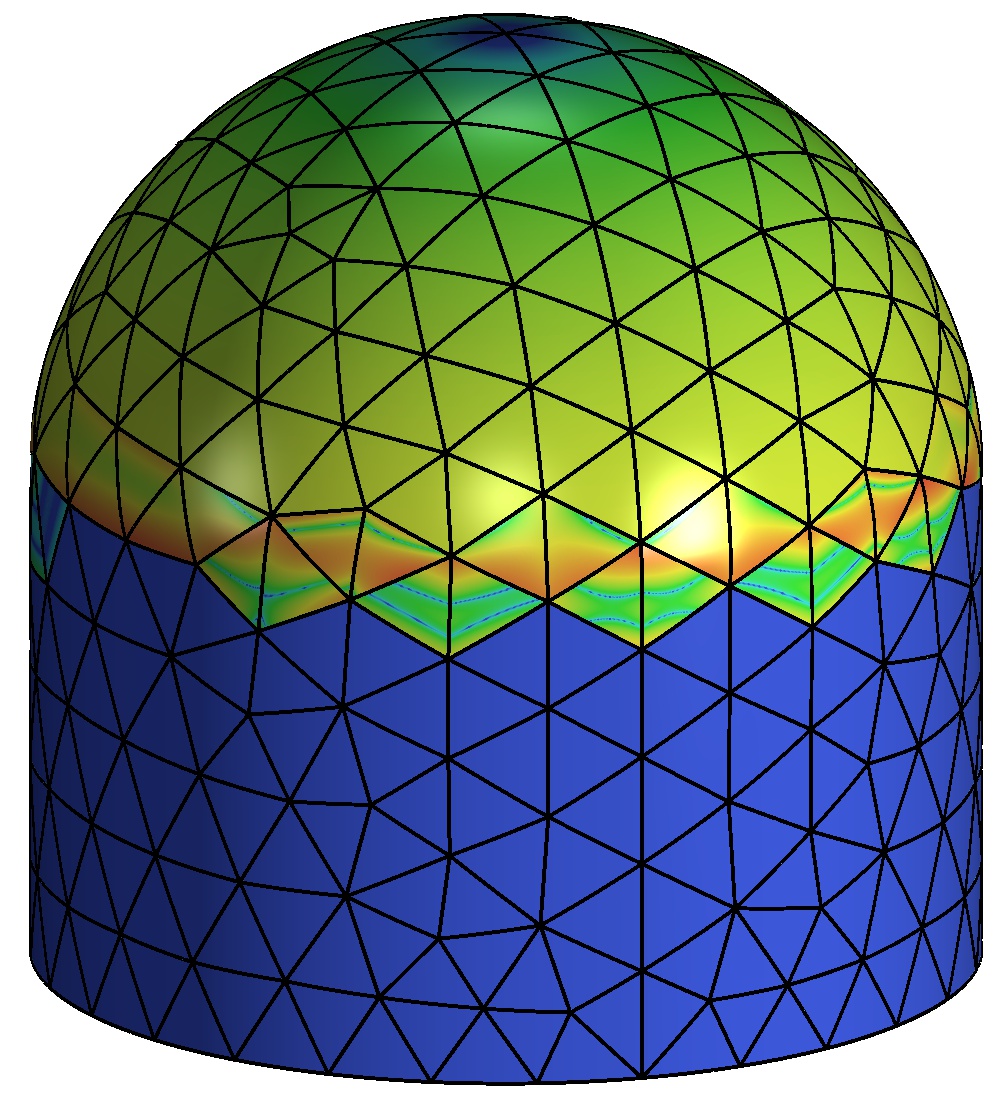}
		\caption{}
		\label{fig:bullet_bad_gradient}
	\end{subfigure}
	\begin{subfigure}[b]{0.08\textwidth}
		\includegraphics[width=\textwidth]{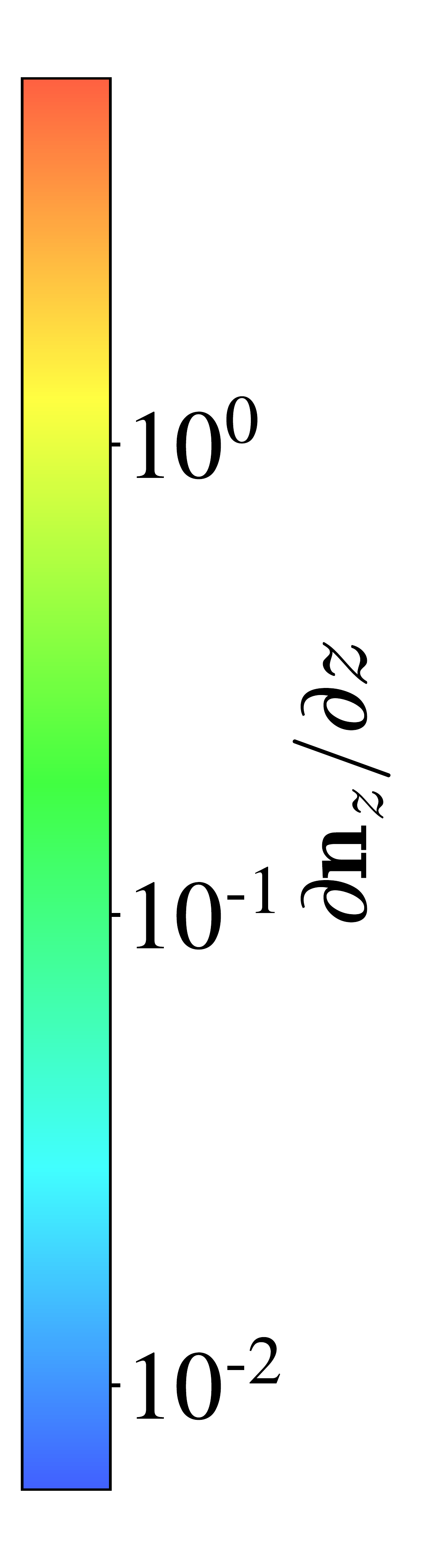}
	\end{subfigure}
	\hfill\hspace{0cm}
	\caption{CAD model colored according to the different boundary entities using:
		(a) the original CAD surfaces; and
		(b) a virtual model.}
	\label{fig:bullet_bad}
\end{figure*}

\subsubsection{Requirements of the virtual geometry}

The process of creating a virtual model has to take into account the continuity of the normal vector in the virtual surfaces. That is, the original surfaces contained in a virtual surface have to define a patch with a continuous normal vector. Otherwise, the final curved mesh may contain oscillations in the boundary approximation. To illustrate this issue, we generate a curved mesh for a virtual model, see Figure \ref{fig:bullet_bad_CAD}, in which the upper sphere and the lower cylinder do not meet with $G^1$ continuity. In particular, the angle between the surfaces normal is seven degrees. The curved mesh of polynomial degree four, Figure \ref{fig:bullet_bad_mesh}, presents oscillations around the normal discontinuity. This is so since the curved triangles have a continuous normal and try to approximate a geometry with a discontinuous normal. In the rest of the domain, the geometry is smooth enough and so is the mesh. We show the derivative of the $z$ component of the normal vector along the $z$ direction in Figure \ref{fig:bullet_bad_gradient}. The elements fully contained either on the cylinder or the sphere present a smooth gradient of the normal vector. However, the elements contained on the interface present large values and high oscillations of the normal vector gradient.

\subsubsection{Requirements of the linear mesh}

With our mesh curving method, we can generate the required meshes for the high-lift common research model. Nevertheless, to accomplish this, we need a linear mesh that satisfies several the curving requirements. Specifically, the linear mesh cannot contain element configurations that hampers the generation of the curved mesh. This leads non-linear problems with high condition numbers and to invalid curved elements. That is, we have non-linear problems that are difficult to solve and, even if we solve them, we obtain an invalid curved mesh. To solve this issue, we could detect the invalid element configurations and perform splitting and flipping operations to obtain a valid linear mesh.

Our mesh curving tools are mature enough to curve a single mesh in minutes. From our perspective, the most difficult part of the process is to obtain a valid linear mesh that can be curved. To perform this task, we may spend several days repairing the linear mesh and removing artifacts that impede the curving process.

\subsubsection{Computational time and memory bottlenecks}

The main time bottleneck to obtain a curved high-order mesh is the definition of an appropriate virtual model and the generation of a linear mesh. On the one hand, the time spent in these two steps is measured in days. On the other hand, our distributed parallel implementation executed with 768 processors is able to curve a single mesh in minutes. This is important since the process of generating a curved mesh is iterative. That is, the generation of a curved mesh allows us to detect artifacts that we need to correct in the virtual model and the linear mesh. The new virtual model and linear mesh are used to create an improved curved mesh to further detect additional artifacts. Thus, the ability to curve a mesh in a small amount of time is essential in the iterative process of mesh generation.

Our main memory bottleneck that limits the largest mesh that we can curve is the generation of the linear mesh. This is the only part of the algorithm that we perform sequentially. Although we can use workstations with more memory, we would ideally require a distributed parallel linear mesh generator.

\subsubsection{Integration of the different programs}

In our mesh curving methodology, we use several programs to repair the CAD model, generate the linear mesh, curve the linear mesh and perform the visual inspection. Nevertheless, these programs are not integrated into a single mesh generation tool. In addition, there are some steps of the curving process that we perform manually. For instance, we need to classify the boundary integer identifiers into the far field, symmetry plane and wall boundary conditions. At this point, we manually perform this task.

\subsection{What mesh quality metrics are used to evaluate high order meshes?}

We have used two types of quality metrics to evaluate a curved mesh. The first one are the element quality metrics that check the elements of the mesh, and the second ones are the geometric accuracy measures that check the approximation properties of the mesh.

\subsubsection{Element quality metrics}

We have used two quality metrics to assess the mesh quality: the scaled Jacobian and the shape quality measure. Both quality metrics detect the inverted elements and therefore, we can check the mesh validity with these quality metrics. Moreover, the element quality metrics also check the quality of the element. In particular, the scaled Jacobian measures how curved is an element, while the distortion quality measures how deviated is the curved element respect to the linear one.

Although the mesh validity is required to perform a simulation, the mesh quality has also an important role in the quality and robustness of the simulation. Low quality elements may introduce spurious artifacts in the numerical solution, and may hamper the convergence of the non-linear problems. For this reason, we advocate to converge the curving problem with tight tolerances. Thus, we avoid oscillations in the element shape that lead to low-quality elements.

\subsubsection{Geometric accuracy metrics}

We have used three different geometric accuracy measures: the maximum point-wise distance, $d_\infty$, an average of the point-wise distance, $SC$, and an  $\mathcal L_2$-average of the point-wise distance, $d_2$. The maximum of the point-wise distance informs about the worst approximation triangle. This is useful to detect which triangles have to be refined to improve the geometric accuracy. The $\mathcal L_2$-average of point-wise distance penalizes the higher distances values and therefore, leads to higher values than $SC$.

Even if the mesh is geometrically accurate, it may contain oscillations that are not detected by the geometric accuracy measures. These oscillations may introduce spurious artifacts in the numerical simulation since the boundary is not smooth. In our methodology, we detect the oscillations of the boundary triangles during the visual inspection of the mesh.

\subsection{How well do the curved meshes conform to the actual geometry?}

For the presented meshes, the maximum point-wise distance is seven orders of magnitude smaller than the aircraft length. The $\mathcal L_2$-average distance ranges between six and seven orders of magnitude smaller than the aircraft length. Finally, the average of the point-wise distance ranges between four and five orders of magnitude smaller than the aircraft length.

For a fixed linear mesh, as we increase the polynomial degree, the resulting curved meshes become more accurate. This is expected since higher polynomial degrees lead to richer approximation spaces and therefore, the curved mesh can better approximate the target geometry.

The approximation properties of the curved mesh mainly depend on the curved boundary triangles. As shown in Table \ref{tab:distances}, the geometric accuracy metrics remain mostly constant with the different $Y+$ values. Thus, we can interpret the target geometry as a boundary condition that the curved mesh has to approximate.

We need a linear mesh that leads to a curved mesh with sufficient geometric accuracy. Mesh generation tools can predict an element size to bound the approximation error of a linear mesh. Nevertheless, this prediction of the element size is not accurate to generate a curved mesh. As a consequence, we cannot check \emph{a priori} if we the prescribed element size is valid. For this reason, we have to iterate the process of element size prescription until we obtain the desired geometric accuracy.

\section{Concluding remarks}
\label{sec:concludingRemarks}

In our experience, the pre-process time is orders of magnitude larger than the curving time. Thus, we need to improve the efficiency and the automation of the pre-process step. In particular, we require tools that help defining the virtual model and prepare the linear meshes. In the specific case of mesh curving for the CRM-HL, preparing curving-friendly inputs took days of human work, while the curving step took minutes of computational time.

\section*{Acknowledgments}

This project has received funding from the European Research Council (ERC) under the European Union's Horizon 2020 research and innovation programme under grant agreement No 715546. This work has also received funding from the Generalitat de Catalunya under grant number 2017 SGR 1731. The work of Xevi Roca has been partially supported by the Spanish Ministerio de Econom\'ia y Competitividad under the personal grant agreement RYC-2015-01633. We acknowledge PRACE for awarding us access to MareNostrum at Barcelona Supercomputing Center (BSC), Spain.

\bibliographystyle{unsrt}
\bibliography{bibliography}

\end{document}